\documentclass[a4paper,12pt]{article}
\usepackage[all]{xy}
\usepackage{amsfonts}
\begin{document}
\title{A Game Theoretic Approach\\ to Quantum Information}
\author{Xianhua Dai  and V. P. Belavkin \\
School of Mathematical Sciences\\
University of Nottingham, UK}

\date{}

\maketitle

\abstract{This work is an application of game theory to quantum information. In a state estimate, we are given observations distributed according to an unknown distribution $P_{\theta}$ (associated with award $Q$), which Nature chooses at random from the set $\{P_{\theta}: \theta \in \Theta \}$ according to a known prior distribution $\mu$ on $\Theta$, we produce an estimate $M$ for the unknown distribution $P_{\theta}$, and in the end, we will suffer a relative entropy cost $\mathcal{R}(P;M)$, measuring the quality of this estimate, therefore the whole utility is taken as $P \cdot Q -\mathcal{R}(P; M)$.

In an introduction to strategic game, a sufficient condition for minimax theorem is obtained; An estimate is explored in the frame of game theory, and in the view of convex conjugate, we reach one new approach to quantum relative entropy, correspondingly quantum mutual entropy, and quantum channel capacity, which are more general, in the sense, without Radon-Nikodym (RN) derivatives. Also the monotonicity of quantum relative entropy and the additivity of quantum channel capacity are investigated.}

\section{Introduction}
Much of quantum information has been concerned with the estimate. We are given observations distributed according to an unknown distribution $P_{\theta}$ (associated with award $Q$), which Nature chooses at random from the set $\{P_{\theta}: \theta \in \Theta \}$ according to a known prior distribution $\mu$ on $\Theta$, we produce an estimate $M$ for the unknown distribution $P_{\theta}$. In the end, we will suffer a cost, measuring the quality of this estimate, therefore the whole utility is in terms of award and cost. One such cost function is relative entropy function $\mathcal{R}(P;M)$, important in several fields, such as information theory, data compression, computational learning theory, game theory, statistics, statistical mechanics, and econometrics.

In the source coding interpretation of the estimate, the minimax value of this game can be interpreted as the capacity of the channel from $\Theta$ to $X$; In computational learning theory, the minimax value of this game is the utility shared by an adaptive algorithm, predicting each observation before it arrives on the previous observation, compared to an algorithm predicting after knowing the real distribution; In gambling theory and mathematical finance, the relative entropy measures the expected reduction in the logarithm of compounded wealth due to lack of knowledge of the true distribution, thus the minimax value of this game is the practical compounded wealth.

In this paper, an introduction to strategic game will be briefly given, during which a sufficient condition for minimax theorem is obtained; an estimate is explored in the frame of game theory, and in the view of convex conjugate, we reach one new approach to quantum relative entropy, quantum mutual entropy, and quantum channel capacity, which are more general, in the sense, without Radon-Nikodym (RN) derivatives. Also the monotonicity of quantum relative entropy and the additivity of quantum channel capacity will be discussed.

The structure of the paper is organized as follows: In the second section, we will give a brief introduction to strategic game, during which a sufficient condition for minimax theorem is obtained; In the third section, we will introduce convex conjugate along with some examples, mainly on its important mathematical properties for our application; In the fourth section, we introduce an estimate in the frame of game theory with the cost of classical relative entropy and reach one new approach to classical relative entropy in the view of convex conjugate; Similarly, we introduce this approach to quantum relative entropy, in the section five, especially the monotonicity of quantum relative entropy is discussed, and further one approach to quantum mutual entropy will be given in the sixth section; The section seven is for quantum channel capacity and its additivity; Final section is for conclusion and possible further problems.

\section{Strategic Game}
Extending the simpler optimization approach developed in variation methods and mathematical programming in mathematics, optimization theory and algorithms in information and computer science, and operation research in neoclassical economics, game theory studies situations where multiple players make decisions in an attempt to maximize their returns. The essential feature is to provide a formal modeling approach to social or informational situations in which decision makers interact with other agents.

Some game theoretic analysis appear similar to decision theory, but game theory studies decisions made in an environment in which players interact. Alternatively, game theory studies choice of optimal behavior when costs and benefits of each option depend upon the choices of other individuals. In this sense, game theory, much more than decision theory, is of similar spirit and situation to information and computer science.

Since Von Neumann and and O. Morgenstern's classic Theory of Games and Economic Behavior [NM 1944] in 1944, there are many introductions to game theory, such as [OR 1994]. In this section, we briefly introduce strategic game, mainly general definitions, existence theorems, and competitive game.

\subsection{General Definitions}

In game theoretic models, the basic entity is a player. A player may be interpreted as an individual or as a group of individuals making a decision. Once we define the set of players, we may distinguish between two types of models: those in which the sets of possible strategies of individual players are primitive; those in which the sets of possible joint strategies of groups of players are primitive. Models of the first type are referred to as "noncooperative".

A strategic game is a model of interactive decision-making in which each decision-maker chooses his plan of strategy once for all, and that these choices are made simultaneously. The model consists of a finite set $N$ of players and, for each player $i$, a set of $A_i$ of strategies and a utility function on the set of strategy profiles $A_1\times\ldots \times A_N$.

\textbf{Definition 1} ([Borel 1953], [Neumann 1928])  A non-cooperative finite strategic game consists of
\begin{itemize}
\item
a finite set $N$ ( the set of players)

and for each player $i\in N$,
\item
a set $A_i$ (the set of strategies available to player $i$ on strategy profile)
\item
a payoff function $u_i:A\rightarrow R$, where $A \equiv \times_{j=1}^N A_j$ (the payoff of player $i$).
\end{itemize}

Denote by $\Sigma_i$ the set of probability measures over $A_i$, which are player $i$'s mixed strategies. And denote by the suffix $-i$ all players except $i$.

In the play of a strategic game, each player holds the correct expectation about the other players' behavior and acts rationally, thus a steady state is reached. If not attempting to examine the process by which a steady state is reached, we call it Nash equilibrium.

\textbf{Definition 2} ([Nash 1950]) A mixed Nash equilibrium of a finite strategic game $\langle N,(A_i),(u_i)\rangle$ is a vector $(\pi_1,\pi_2,\ldots,\pi_n)$, with $\pi_i\in\Sigma_i$ for all $i\in N$, such that
\begin{equation}
\Sigma_{a\in A}\pi_i(a_i)\pi_{-i}(a_{-i})u_i(a_i,a_{-i})\geq\Sigma_{a\in A}\rho_i(a_i)\pi_{-i}(a_{-i})u_i(a_i,a_{-i})
\end{equation}

\noindent
for all $\rho_i\in\Sigma_i$ and for all $i\in N$.

Therefore, pure-strategy Nash equilibria are those which only involve degenerate mixed strategies.

The following restatement of the definition is useful elsewhere.

\textbf{Definition} For any $a_{-i} \in A_{-i}$, we define $B_i(a_{-i})$ best actions to be the set of player $i'$s given $a_{-i}$:
\begin{equation}
B_i(a_{-i})= \{ a_i \in A_i: u_i(a_i,a_{-i}) \geq u_i(a_i^\prime ,a_{-i}) \}.
\end{equation}

\noindent
The set-valued function $B_i$ is called the best-response function of player $i$. Therefore, a Nash equlibrium is a profile $a^*$ of actions for which
\begin{equation}
a_i \in B_i(a_{-i}^*),
\end{equation}

\noindent
for all $i \in N$.

This alternative definition formulation points us to a (not necessarily efficient) method of finding Nash equilibria: At first to calculate the best response function of each player, then to find a profile $a^*$ of actions for which $a^* \in B_i(a_{-i}^*)$ for all $i \in N$. Obviously, if the function $B_i$ are singleton-valued, the second step deduces to solve $|N|$ equations in the $|N|$ unknowns $(a^*_i)_{i \in N}$.

\subsection{Existence Theorems}

An existence result has two purposes: At first, if we have a game satisfying the hypothesis of the result, it is hopeful to find an equilibrium; Secondly, the existence of an equilibrium ensures the game consistent with a steady state solution; Furthermore, the existence of an equilibria for a family of games allows us to study properties of these equilibria without finding them explicitly and without the risk to study the empty set.

It is extensively investigated that under which conditions the set of Nash equilibria of a game is nonempty. We here just introduce one of the simplest existence result, whose mathematical level is much more advanced.

To prove that a Nash equilibrium exists for a game, it suffices to show that there is a profile $a^*$ of actions such that $a^* \in B_i(a_{-i}^*)$ for all $i \in N$, which is $a^* \in B(a^*)$, if we define the set-valued function $B: A \longrightarrow A$ by $B(a)= \times_{i \in N} B_i(a_{-i})$. Luckily fixed point theorems give conditions on $B$ under which there exists a value of $a^*$ for which $a^* \in B(a^*)$. Generally we apply the following fixed point theorem.

\textbf{Theorem} ([Kakutani 1941]) Let $X$ be a compact convex subset of $\mathbb{R}^n$ and let $f: X \longrightarrow X$ be a set-valued function for which
\begin{itemize}
\item
for all $x \in X$, the set $f(x)$ is nonempty and convex; and
\item
the graph of $f$ is closed (i.e. for all sequences $\{ x_n \}$ and $\{ y_n \}$ such that $\{ y_n \} \in f(\{ x_n \})$ for all $n$, $x_n \longrightarrow x$ and $y_n \longrightarrow y$, we have $y \in f(\{ x \})$).
\end{itemize}

\noindent
Then there exists a $x^* \in f(x^*)$.

\textbf{Theorem} ([Nash 1950, 1951]) A Nash equilibrium of strategic game $\langle N,(A_i),(u_i)\rangle$ exists if for all $i\in N$,
\begin{itemize}
\item
the set $A_i$ of actions of player $i$ is a nonempty compact convex subset of a Euclidian space; and
\item
the utility function $u_i$ is continuous and quasi-concave on $A_i$.
\end{itemize}

Proof: Let set-valued function $B_i$ the best-response function of player $i$, we define $B: A \longrightarrow A$ by $B(a)= \times_{i \in N} B_i(a_{-i})$.

For every $i\in N$, the set $B_i(a_{-i})$ is nonempty since the utility function $u_i$ is continuous and the set $A_i$ is compact, and also is convex since the utility function $u_i$ is quasi-concave on $A_i$; $B$ has a closed graph since each utility function $u_i$ is continuous.

Following the Kakutani's fixed point theorem, $B$ has a fixed point; any fixed point is a Nash equilibrium of the game as noted.

This result states that a strategic game satisfying certain conditions has at least one Nash equilibrium; but a game may have more than one equilibrium. Note that this theorem does not apply to any game in which some player has finitely many actions, since the set of actions of every player is not convex, but a mixed strategy Nash equilibrium of every finite strategic game $\langle N,(A_i),(u_i)\rangle$ always exists.

\textbf{Theorem} ([Nash 1950]) A mixed strategy Nash equilibrium of every finite strategic game $\langle N,(A_i),(u_i)\rangle$ always exists.

Proof: Let $G=\langle N,(A_i),(u_i)\rangle$ be a strategic game, and for each player $i$, let $m_i$ be the number of members of the set $A_i$, then we identify the set $\Delta (A_i)$ of player $i'$s mixed strategies with the set of vectors $(p_1,p_2,...,p_{m_i})$ for which $p_k \geq 0$ for all $k$ and $ \sum_{k=1}^{m_i} p_k=1$ ($p_k$ being the probability with which player $i$ uses his $i$th pure strategy). This set is nonempty, convex, and compact. Since expected payoff is linear in the probabilities, each player's payoff function in the mixed extension of $G$ is quasi-concave in his strategy and continuous. Thus a mixed strategy Nash equilibrium exists due to above theorem.

\subsection{Competitive Game}

Little obtained on the set of Nash equilibria of an arbitrary strategic game, we discuss strictly competitive games and its qualitative character of the equilibria.

\textbf{Definition} A strategic game $\langle \{1,2\},(A_i),(u_i)\rangle$ is strictly competitive if for any $a \in A$ and $b \in A$, we have $u_1(a)\geq u_1(b)$ if and only if $u_2(a) \leq u_1(b)$.

Player $i$ maxminimizes if he chooses an action best for him on the assumption that whatever he does, player $j$ will choose her action to hurt him as much as possible. We will find that for a strictly completive game possessing a Nash equilibrium, a pair of actions is a Nash equilibrium if and only of the action of each player is a maxminimizer, a striking result since providing a link between individual decision-making and the reasoning behind the notion of Nash equilibrium, during which we also find that for a strictly completive game possessing Nash equilibria yield the same payoffs.

\textbf{Definition} Let $G=\langle \{1,2\},(A_i),(u_i)\rangle$ be a strictly completive game, the action $x^* \in A_1$ is a maxminimizer for player 1 if for all $x \in A_1$,
\begin{equation}
\min_{y \in A_2}u_1(x^*,y) \geq \min_{y \in A_2} u_1(x,y).
\end{equation}

\noindent
Similarly, the action $y^* \in A_2$ is a maxminimizer for player 2 if for all $y \in A_2$,
\begin{equation}
\min_{x \in A_1}u_2(x,y^*) \geq \min_{x \in A_1} u_2(x,y).
\end{equation}

\textbf{Theorem} ([Neumann 1928]) Let $G=\langle \{1,2\},(A_i),(u_i)\rangle$ be a strictly completive game.

a) If $(x^*,y^*)$ is a Nash equilibrium of $G$, then $x^*$ is a maxminimizer for player 1 and $y^*$ is a maxminimizer for player 2.

b) If $(x^*,y^*)$ is a Nash equilibrium of $G$, then
\begin{equation}
\max_x \min_y  u_1(x,y) = \min_y \max_x  u_1(x,y) = u_1(x^*,y^*),
\end{equation}

\noindent
and all Nash equilibria of $G$ yield the same payoffs.

c) If $\max_x \min_y  u_1(x,y) = \min_y \max_x  u_1(x,y)$, $x^*$ is a maxminimizer for player 1, and $y^*$ is a maxminimizer for player 2, then $(x^*,y^*)$ is a Nash equilibrium of $G$.

Proof: First to prove (a) and (b). Let $(x^*,y^*)$ is a Nash equilibrium of $G$, then $u_2(x^*,y^*) \geq u_2(x^*,y)$ for all $y \in A_2$, or $u_1(x^*,y^*) \leq u_1(x^*,y)$ for all $y \in A_2$. Hence
\begin{equation}
u_1(x^*,y^*) = \min_y  u_1(x^*,y) \leq \max_x \min_y  u_1(x,y).
\end{equation}

Similarly,
\begin{equation}
u_1(x^*,y^*)\geq \max_x \min_y  u_1(x,y).
\end{equation}

Thus $u_1(x^*,y^*) = \max_x \min_y  u_1(x,y)$ and $x^*$ is a maxminimizer for player 1. Similar argument for player 2, $y^*$ is a maxminimizer for player 2 and $u_2(x^*,y^*) = \max_y \min_x  u_2(x,y)$.

Now to prove (c). Let $v^* = \max_x \min_y  u_1(x,y) = \min_y \max_x  u_1(x,y)$, for a strictly completive game, we have $-v^* = \max_y \min_x  u_2(x,y)$. Since $x^*$ is a maxminimizer for player 1, we have $u_1(x^*,y) \geq v^*$ for all $y \in A_2$; similarly, $u_2(x,y^*) \geq - v^*$ for all $x \in A_1$. Taking $y=y^*$ and $x=x^*$ in those two inequalities, we have $v^* =u_1(x^*,y^*)$, again considering this strictly completive game, we reach that $(x^*,y^*)$ is a Nash equilibrium of $G$.

Following part (c), a Nash equilibrium can be found by solving the problem
\begin{equation}
\max_x \min_y  u_1(x,y);
\end{equation}

\noindent
following part (a) and (c), Nash equilibria of strictly completive game are interchangeable: if $(x,y)$ and $(x^{\prime},y^{\prime})$ are equilibria, so are $(x,y^{\prime})$ and $(x^{\prime},y)$; following (b), for any strictly competitive game with a Nash equilibrium,
\begin{equation}
\max_x \min_y  u_1(x,y) = \min_y \max_x  u_1(x,y) = u_1(x^*,y^*).
\end{equation}

If $\max_x \min_y  u_1(x,y) = \min_y \max_x  u_1(x,y) = u_1(x^*,y^*)$, we say this equilibrium payoff of player 1 is the value of the game.

\textbf{Theorem} Let $A_1$, $A_2$ be non-empty, convex and compact subsets of $\mathbb{R}^n$ for some $n$. Let payoff $u:A_1 \times A_2 \longrightarrow \mathbb{R}$ be a continuous function, such that
\begin{itemize}
\item
$\forall a_2 \in A_2$, the set $ \{ a_1 \in A_1: u(a_1,a_2) \geq u(a_1^\prime,a_2), \forall a_1^\prime \in A_1 \}$ is convex; and

\item
$\forall a_1 \in A_1$, the set $ \{ a_2 \in A_2: u(a_1,a_2) \leq u(a_1,a_2^\prime), \forall a_2^\prime \in A_2 \}$ is convex.
\end{itemize}

\noindent
Then there exists an $a^* \in A_1 \times A_2$, such that
\begin{equation}
\max_{a_1 \in A_1}\min_{a_2 \in A_2} u(a_1,a_2)=u(a^*)=\min_{a_2 \in A_2}\max_{a_1 \in A_1} u(a_1,a_2).
\end{equation}

Proof: At first, continuous payoff function $u$ is quasi-concave with respect to two arguments, since $\forall a_2 \in A_2$, the set $ \{ a_1 \in A_1: u(a_1,a_2) \geq u(a_1^\prime,a_2), \forall a_1^\prime \in A_1 \}$ is convex, and $\forall a_1 \in A_1$, the set $ \{ a_2 \in A_2: u(a_1,a_2) \leq u(a_1,a_2^\prime), \forall a_2^\prime \in A_2 \}$ is convex.

Following [Nash 1950, 1951], a Nash equilibrium $a^* \in A_1 \times A_2$ of this strategic game exists.

Further according to [Neumann 1928], for this competitive game,
\begin{equation}
\max_x \min_y  u(x,y) = \min_y \max_x  u(x,y) = u(a^*).
\end{equation}

\section{Convex Conjugate}

In mathematics, convex conjugation, as a generalization of the Legendre transformation (in which sense, is also taken as Legendre-Fenchel transformation or Fenchel transformation elsewhere), addressed much attention in the study of extremum problems, among which are system inequalities, the minimum or maximum of a convex function over a convex set, Lagrange multipliers, and minimax theorems.

There are excellent books on the introduction to convex analysis, such as [Rockafellar 1970] for pure mathematics, [Arnold 1989] for application in theoretic mechanics. In this section, we simply overview convex conjugation, first on its definition including some examples, then on some properties for our application.

\subsection{General Definition}
\textbf{Definition} Let $X$ be a linear normed space, and $X^*$ the dual space to $X$, we denote the dual pairing by
\begin{equation}
\langle . , . \rangle: X^* \times X \longrightarrow \mathbb{R}.
\end{equation}

\noindent
Given a function $f: X \longrightarrow \mathbb{R} \cup \{ +\infty \}$ taking values on the extended real number line, we define the convex conjugate
$f^*: X^* \longrightarrow \mathbb{R} \cup \{ +\infty \}$ by
\begin{equation}
f^*(x^*) \equiv \sup \{\langle x^*,x \rangle -f(x) | x \in X \},
\end{equation}

\noindent
or, equivalently, by
\begin{equation}
f^*(x^*) \equiv - \inf \{f(x) - \langle x^*,x \rangle | x \in X \}.
\end{equation}

We consider convex conjugates for some examples, via simple computations, following the above definition.

\textbf{Example 1} An affine function is generally written by
\begin{equation}
f(x) \equiv \langle a,x \rangle -b, a \in \mathbb{R}^n, b \in \mathbb{R}.
\end{equation}

\noindent
Then its convex conjugate $f^*(x^*)$, denoted by $O_a(x^*)$, is
$$
f^*(x^*) = O_a (x^*) = \left\{
\begin{array}{ll}
a \ \ \ \ x^*=a; \\
\infty \ \ \ \ x^* \neq a. \\
\end{array}
\right.
$$

\textbf{Example 2} The norm function is generally written by
\begin{equation}
f(x) \equiv \|x\|.
\end{equation}

\noindent
Then its convex conjugate $f^*(x^*)$, denoted by $O_1(x^*)$, is
$$
f^*(x^*) \equiv O_1 (x^*) = \left\{
\begin{array}{ll}
a \ \ \ \ \| x^* \| \leq 1; \\
\infty \ \ \ \ \| x^* \| = 1. \\
\end{array}
\right.
$$

\textbf{Example 3} Let $K \subseteq X$ be a convex subset and $e(x)$ be the calibration function
\begin{equation}
e(x) \equiv \sup \{\langle x^*,x \rangle | x^* \in K \}.
\end{equation}

Then its convex conjugate $e^*(x^*)$ is $O_K(x^*)$, where $O_K(x^*)$ is defined as follows.
$$
e^*(x^*) \equiv O_K(x^*) = \left\{
\begin{array}{ll}
0 \ \ \ \ x^* \in K; \\
\infty \ \ \ \ x^* \in K^c, \\
\end{array}
\right.
$$

\noindent
where $K^c$ is the complement of $K$.

\textbf{Example 4} The convex conjugate $f^*(x^*)$ of exponential function $f(x)=e^x$ is
$$
f^*(x^*) = \left\{
\begin{array}{ll}
x^* \ln x^* -x^* \ \ \ \ x^*>0; \\
0 \ \ \ \ x^*=0; \\
\infty \ \ \ \ x^* < 0. \\
\end{array}
\right.
$$

\noindent
Let a cone $X_+$ be $\{x \geq 0 \} \subseteq X$, and $X^*_+ \equiv \{x^* \in X^*: \langle x^*,x \rangle \geq 0 \}$ its dual cone. Then the convex conjugate $f^*(x^*)$ of exponential function $f(x)=e^x$ on $X_+$ is
$$
f^*(x^*) = \left\{
\begin{array}{ll}
x^* \ln x^* -x^* \ \ \ \ x^*>0; \\
0 \ \ \ \ x^*=0. \\
\end{array}
\right.
$$

\subsection{Some Properties}
\textbf{Theorem} The conjugate function of a closed convex function is a closed convex function.

Proof: For every $t \in \mathbb{R} \bigcap [0,1]$, and every $x^*,y^* \in X^*$, according to the definition of convex conjugate,
\begin{equation}
f^*(tx^*+(1-t)y^*) \equiv \sup \{\langle tx^*+(1-t)y^*,x \rangle -f(x) | x \in X \}
\end{equation}
\begin{equation}
= \sup \{(\langle tx^*,x \rangle -t f(x))+(\langle (1-t)y^*,x \rangle -(1-t) f(x)) | x \in X \}
\end{equation}
\begin{equation}
\leq  t \sup \{ \langle x^*,x \rangle - f(x)| x \in X \}+ (1-t) \sup \{\langle y^*,x \rangle -f(x) | x \in X \}
\end{equation}
\begin{equation}
\equiv tf^*(x^*)+ (1-t)f^*(y^*).
\end{equation}

\textbf{Theorem} (Order-reversing) Convex-conjugation is order-reversing, i.e., if $f \leq g$, then $f^* \geq g^*$, where $f \leq g$ means for every $x \in X$, $f(x) \leq g(x)$.

Proof: Since $f \leq g$, then for every $x \in X$,
\begin{equation}
f(x) \leq g(x).
\end{equation}

According to the definition of convex conjugate, for every $x^* \in X^*$,
\begin{equation}
f^*(x^*) \equiv \sup \{\langle x^*,x \rangle -f(x) | x \in X \}
\end{equation}
\begin{equation}
\geq \sup \{\langle x^*,x \rangle -g(x) | x \in X \}
\end{equation}
\begin{equation}
=g^*(x^*).
\end{equation}

\noindent
Thus $f^* \geq g^*$, since every $x^* \in X^*$.

\textbf{Theorem} (Biconjugate) The convex conjugate of a function is lower semi-continuous. The biconjugate $f^{**}$ (the convex conjugate of the convex conjugate) is the closed convex hull, that is, the largest lower semi-continuous convex function smaller than $f$. Furthermore, for proper functions $f$, $f = f^{**}$ if and only if $f$ is convex and lower semi-continuous.

Proof: For every $x^* \leq x_0^* \in X^*$,
\begin{equation}
f^*(x^*) \equiv \sup \{\langle x^*,x \rangle -f(x) | x \in X \}
\end{equation}
\begin{equation}
\leq \sup \{\langle x_0^*,x \rangle -f(x) | x \in X \}
\end{equation}
\begin{equation}
\equiv f^*(x_0^*),
\end{equation}

\noindent
which implies that the convex conjugate of a function is lower semi-continuous.

\textbf{Theorem} (Fenchel's inequality or Fenchel-Young inequality) For any proper convex function $f$ and its convex conjugate $f^*$, Fenchel's inequality holds:
\begin{equation}
\langle p,x \rangle \leq f(x)+ f^*(p),
\end{equation}

\noindent
for every $x \in X, p \in X^*$.

Proof: According to the definition of convex conjugate, for every $x \in X, p \in X^*$,
\begin{equation}
f(x)+ f^*(p)=f(x)+ \sup \{\langle p,x \rangle -f(x) | x \in X \}
\end{equation}
\begin{equation}
\geq f(x)+ (\langle p,x \rangle -f(x))
\end{equation}
\begin{equation}
=\langle p,x \rangle.
\end{equation}

\textbf{Theorem} (Infimal convolution) Let $f_1,..., f_m$ be proper convex functions on $X$. Then
\begin{equation}
(f_1 \star_{\inf}...\star_{\inf} f_m)^*=f_1 ^*+...+ f_m^*,
\end{equation}

\noindent
where the infimal convolution of two functions $f$ and $g$ on $X$ is defined as
\begin{equation}
(f \star_{\inf} g)(x) \equiv \inf \{ f(x-y)+ g(y) | y \in X \}.
\end{equation}

Proof: Here we just consider the case for $m=2$, for $x \in X$
\begin{equation}
(f\star_{\inf}g)^*(x^*)=(\inf \{ f(x-y)+ g(y) | y \in X)^*(x^*)
\end{equation}
\begin{equation}
=\sup \{\langle x^*,x \rangle -\inf \{ f(x-y)+ g(y) | y \in X \}| x \in X \}
\end{equation}
\begin{equation}
= \sup \{\langle x^*,(x-y)+y \rangle -f(x-y)- g(y) | x, y \in X \}
\end{equation}
\begin{equation}
= \sup \{\langle x^*,x-y \rangle -f(x-y) | x, y \in X \}+ \sup \{\langle x^*,y \rangle -g(y) | y \in X\}
\end{equation}
\begin{equation}
= f^*(x^*)+g^*(x^*);
\end{equation}

Since the infimal convolution is associative, i.e.,
\begin{equation}
[(f_1 \star_{\inf}...\star_{\inf} f_{m-1}) \star_{\inf} f_m]^*=(f_1 \star_{\inf}...\star_{\inf} f_{m-1})^*+ f_m^*,
\end{equation}

\noindent
the theorem follows from mathematical induction for general $m$.

\section{Classical information}
In classical information theory, we need find fundamental limits on compressing and reliably communicating classical data. A key measure of information is known as information entropy, usually expressed by the average number of bits needed for storage or communication.

This section introduces basic classical information quantities, say Shannon entropy, relative entropy, see, for example, [Shannon 1948], for reference. We first overview basic mathematical forms of Shannon entropy and relative entropy, then explore classical estimation in the frame of game theory, and reach a new approach to relative entropy in the view of convex conjugate.

\subsection{Classical Relative Entropy}
Suppose there is a random variable with true distribution $F$ (for the density $f$). Then we could represent that random variable with a code of average length $H(F)$, where Shannon entropy $H(F)$ is expressed in mathematics as follows.

\textbf{Definition} (Shannon Entropy $H(F)$ of the distribution $F$)
\begin{equation}
H(F) \equiv - \int f(x) \log f(x) dx.
\end{equation}

However, due to incomplete information (we do not know $F$ really), we assume $G$ the distribution of the random variables instead. Then the code would need more bits to represent the random variable. The difference, in the number of bits, denoted by $\mathcal{R}(F;G)$, between a "true" probability distribution $F$ and an arbitrary probability distribution $G$ is known as the relative entropy [Shannon 1948], or the Kullback–Leibler divergence, information divergence, information gain in probability theory and information theory, which is expressed in mathematics as follows.

\textbf{Definition} (Relative Entropy $\mathcal{R}(F;G)$ of probability distributions $F$ and $G$)
\begin{equation}
\mathcal{R}(F;G) \equiv \int \log (f/g)d F,
\end{equation}

\noindent
where $f$ and $g$ are the respective densities with respect to any dominating measure.

Though the relative entropy $\mathcal{R}(F;G)$ is often taken as a distance metric, but it is not a true metric, since it is not symmetric between distribution $F$ and $G$.

There may be some interpretations for relative entropy, for example, the relative entropy $\mathcal{R}(F;G)$ may be interpreted as the error exponent for the hypothesis test $F$ versus $G$.

\subsection{Classical Estimate}
In a classical estimate, we are given classical observations distributed according to an unknown distribution $P_{\theta} \in X \in \ell$ associated with award $Q$, which Nature chooses randomly from the set $\{P_{\theta}: \theta \in \Theta \}$ according to a known prior distribution $\mu$ on $\Theta$, we produce an estimate $M$ for the unknown distribution $P_{\theta}$, denoted by $P$ later without notation confusion. In the end, we will suffer a relative entropy cost $\mathcal{R}(P;M)$, measuring the quality of this estimate, thus the whole utility is $P \cdot Q -\mathcal{R}(P; M)$.

In this section, we will investigate the existence of minimax value of this utility, correspondingly its minimax strategy.

We consider the utility $P \cdot Q- \mathcal{R}(P;M)$, then the estimate problem is in fact the following optimization problem
\begin{equation}
\min_{M \geq 0, M \cdot I=1} \max_{P \geq 0, P \cdot I=1} [P \cdot Q- \mathcal{R}(P;M)].
\end{equation}

Considering the convex conjugation $\mathcal{R}_M^*(Q)$ of $\mathcal{R}(P;M)$ with respect to $P$, that is,
\begin{equation}
\max_{P \geq 0, P \cdot I=1} [P \cdot Q- \mathcal{R}(P;M)]=\mathcal{R}_M^*(Q),
\end{equation}

\noindent
we can rewrite the above estimate problem as follows.
\begin{equation}
\min_{M \geq 0, M \cdot I=1} \max_{P \geq 0, P \cdot I=1} [P \cdot Q- \mathcal{R}(P;M)]= \min_{M \geq 0, M \cdot I=1} \mathcal{R}_M^*(Q).
\end{equation}

\textbf{Remark} If we take function $\mathcal{R}_1(P)$ as follows.
$$
\mathcal{R}_1(P) = \left\{
\begin{array}{ll}
\mathcal{R}(P;M) \ \ \ \ P \cdot I=1; \\
\infty \ \ \ \ P \cdot I \neq 1. \\
\end{array}
\right.
$$

\noindent
Then we can write in the following form.
\begin{equation}
\max_{P \geq 0} [P \cdot Q- \mathcal{R}_1(P)]=\mathcal{R}_M^*(Q),
\end{equation}

\noindent
where $\mathcal{R}_1(P)=\mathcal{R}(P;M)+ O_{A_1}(P)$, and the hyperplane $A_1=\{ P: P \cdot I=1\}$.

Applying the Lagrange Theorem, we obtain the following result.
\begin{equation}
\mathcal{R}_M^*(Q)=\max_{P \geq 0, P \cdot I=1} [P \cdot Q- \mathcal{R}_M(P)]=\min_{\lambda} \max_{P \geq 0} [P \cdot Q- \mathcal{R}_1(P)+\lambda (P \cdot I - 1)].
\end{equation}

Since the function $P \cdot Q- \mathcal{R}_1(P)+\lambda (P \cdot I - 1)$ is linear with respect to $\lambda$ and convex with respect to $P$, the $\min$ and $\max$ can be exchanged, i.e.,
\begin{equation}
\min_{\lambda} \max_{P \geq 0} [P \cdot Q- \mathcal{R}_1(P)+\lambda (P \cdot I - 1)]=\max_{P \geq 0} \min_{\lambda} [P \cdot Q- \mathcal{R}_1(P)+\lambda (P \cdot I - 1)],
\end{equation}

\noindent
therefore we obtain the following result.
\begin{equation}
\mathcal{R}_M^*(Q)=\max_{P \geq 0, P \cdot I=1} [P \cdot Q- \mathcal{R}_1(P)]= \max_{P \geq 0} \min_{\lambda} [P \cdot Q- \mathcal{R}_1(P)+\lambda (P \cdot I - 1)].
\end{equation}

\textbf{Remark} If considering this optimization problem, at first, with respect to $\lambda$, that is,
\begin{equation}
\max_{P \geq 0} [P \cdot Q- \mathcal{R}_1(P)+O_{A_1}(P)]= \max_{P \geq 0}  [P \cdot Q- \mathcal{R}_1(P)+\min_{\lambda} \lambda (P \cdot I - 1)],
\end{equation}

\noindent
we obtain the following result.
\begin{equation}
\mathcal{R}_M^*(Q)=\max_{P \geq 0}  [P \cdot Q- \mathcal{R}_1(P)+\min_{\lambda} \lambda (P \cdot I - 1)].
\end{equation}

To consider the optimization problem
\begin{equation}
\min_{\lambda} \max_{P \geq 0} [P \cdot Q- \mathcal{R}_1(P)+\lambda (P \cdot I - 1)],
\end{equation}

\noindent
applying the variational method, it suffices to consider the function
\begin{equation}
U=P \cdot Q- \mathcal{R}_1(P)+\lambda (P \cdot I - 1),
\end{equation}

\noindent
or, equivalently, the function
\begin{equation}
\sum_x P_x(Q_x+\ln \frac{P_x}{M_x}+\lambda)-\lambda,
\end{equation}

\noindent
and we obtain the result.
\begin{equation}
\delta U = \delta P_x (Q_x+ \ln \frac{M_x}{P_x}+ \lambda - 1) + \delta \lambda (\sum_x P_x -1),
\end{equation}

\noindent
from which we obtain the following result.
\begin{equation}
Q_x+ \ln \frac{M_x}{P_x}+ \lambda - 1=0.
\end{equation}

Therefore, we reached following two results.
\begin{equation}
P_x^*=M_x \cdot \exp (Q_x +\lambda -1),
\end{equation}
\begin{equation}
\mathcal{R}_M^*(Q)=P^* \cdot Q -\mathcal{R}_M(P^*)= - \sum_x M_x \exp (Q_x +\lambda -1) (\lambda -1).
\end{equation}

Considering further $P \cdot I = 1$, we obtain the following results.
\begin{equation}
\exp (\lambda -1) = \frac{1}{\sum_x M_x \exp Q_x},
\end{equation}
\begin{equation}
P_x^*=\frac{M_x \cdot \exp Q_x}{\sum_x M_x \exp Q_x},
\end{equation}
\begin{equation}
\mathcal{R}_M^*(Q)=P^* \cdot Q -\mathcal{R}_M(P^*)= \ln (\sum_x M_x \exp Q_x).
\end{equation}

\textbf{Remark} It is easy to see that $\mathcal{R}_M(P^*)=P^* \cdot Q - \mathcal{R}_M^*(Q)$ is the classical relative entropy under the maximal utility, or classical relative capacity under given utility.

Since the set $\{ M: M \geq 0, M \cdot I=1 \}$ is convex, the minimum of function $\mathcal{R}_M(P^*)$ always exists with respect to $M$. Therefore, we reach the following main result.

\textbf{Theorem} The minimax value, associated with the above estimate game, defined by
\begin{equation}
\overline{\mathrm{V}}= \inf_{M \geq 0, M \cdot I=1} \sup_{P \geq 0, P \cdot I=1} [P \cdot Q- \mathcal{R}(P;M)],
\end{equation}

\noindent
makes sense, and so does its minimax strategy.

\textbf{Remark} We can similarly define the maxmin value, associated with the above estimate game, by
\begin{equation}
\underline{\mathrm{V}}= \sup_{P \geq 0, P \cdot I=1} \inf_{M \geq 0, M \cdot I=1} [P \cdot Q- \mathcal{R}(P;M)],
\end{equation}

\noindent
but it is needy to consider if this maxmin value and its maxmin strategy always exist, and further if this maxmin value is the same as the minimax value when this maxmin value always exists.

\subsection{Convex Conjugate View}
In the above, we applied the convex conjugate $\mathcal{R}_M^*(Q)$ of classical relative entropy $\mathcal{R}_M(P)$ with respect to $P$, i.e.,
\begin{equation}
\mathcal{R}_M^*(Q)=\max_{P \geq 0, P \cdot I=1} \{P \cdot Q- \mathcal{R}_M(P)\},
\end{equation}

\noindent
and obtained the following formula
\begin{equation}
\mathcal{R}_M^*(Q)= \ln (\sum_x M_x \exp Q_x),
\end{equation}

\noindent
but starting from the result $\mathcal{R}_M^*(Q)= \ln (\sum_x M_x \exp Q_x)$ and applying the above biconjugate property, we can define the classical relative entropy as follows.

\textbf{Definition} Classical relative entropy $\mathcal{R}(\rho; M)$ of $\rho$ relative to $M$ is defined as
\begin{equation}
\mathcal{R}(\rho; M) \equiv \mathcal{R}_M(\rho)=\max_Q \{<\rho, Q> - \mathcal{R}_M^*(Q)\},
\end{equation}

\noindent
where $\mathcal{R}_M^*(Q)= \ln (\sum_x M_x \exp Q_x)$.

Obviously, the simple computation will give us the mathematical form of classical relative entropy.

\textbf{Proposition} Classical relative entropy of $\{\rho_x\}$ relative to $\{M_x\}$ is equal to $\sum_x [\rho_x \log (\rho_x/M_x)]$, that is,
\begin{equation}
\mathcal{R}(\rho; M) =\sum_x [\rho_x \log (\rho_x/M_x)],
\end{equation}

\noindent
which confirms the unique mathematical form of classical relative entropy, though still open to explain $\mathcal{R}_M^*(Q)= \ln (\sum_x M_x \exp Q_x)$ completely in information theory.

\section{Quantum Relative Entropy}
Many information measures for Quantum Signals, for example, Von Newmann Entropy [Wehrl 1978, OP 1993], Quantum Conditional Entropy [HOW 2005], Quantum Relative Entropy [Umegaki 1962], Quantum Mutual Entropy [NC 2000], etc, upon information theoretic explanation and mathematical formula of Von Newmann Entropy. In fact, at present stage, no original information-theoretic definition, similar to shannon information entropy, is possible even for Von Newmann Entropy except for the mathematical formula.

This section is for one new and general quantum relative entropy, in the sense without Radon-Nikodym (RN) derivatives. First we simply overview three types of quantum relative entropy, then investigate quantum estimate in the frame of game theory, (here following the terminology of classical estimate, the same terminology of estimate is still used, but in a different sense away from quantum physics), and reach one new mathematical form of quantum relative entropy, also give some important properties including monotonicity of quantum relative entropy.

Throughout this paper we prefer to use in what is following the term "information" for negaentropy, leaving the term "entropy" for the opposite quantities like relative negainformation $\mathcal{S}(\varpi; \varphi)=-\mathcal{R}(\varpi; \varphi)$, which coincides with usual von Newmann entropy $\mathcal{S}(\varpi)$ if it is taken with respect to the trace $\varphi=\mathrm{Tr}$.

\subsection{Historic Review}
There are several mathematical quantum relative entropies so far, mainly Araki-Umegaki type [Araki 1976], Belavkin-Staszewski type, Hammersley-Belavkin type [HB 2006]. This part overviews the basic definition of quantum relative entropy of Hammersley-Belavkin type and reduces to Araki-Umegaki type, Belavkin-Staszewski type.

We define the Radon-Nikodym (RN) derivatives with respect to an arbitrary, not necessarily normalized, density $\gamma \in \mathcal{A}_\top$ (i.e. a positive linear functional on $\mathcal{A}$),
\begin{equation}
\varrho_{\gamma}=\tilde{\gamma}^{-1/2} \varrho \tilde{\gamma}^{-1/2},
\end{equation}
\begin{equation}
\varsigma_{\gamma}=\tilde{\gamma}^{-1/2} \varsigma \tilde{\gamma}^{-1/2},
\end{equation}

\noindent
then we obtain quantum relative entropy of Hammersley-Belavkin type (known as $\gamma$ type elsewhere) as follows:

\textbf{Definition} Quantum relative entropy of Hammersley-Belavkin type (known as $\gamma$ type elsewhere) to compound state $\varpi$ on the algebra $\mathcal{A} \otimes\mathcal{B}$, (or information divergence of the state $\varpi$ with respect to a reference state $\varphi$) is defined by the density operator $\omega, \phi$ of these states $\varpi$ and $\varphi$ as
\begin{equation}
\mathcal{R}^{g}_{\gamma}(\varpi; \varphi)=\lambda (\sqrt{\varsigma} g(\mathrm{L}_{\varsigma_{\gamma}}^{-1}\mathrm{L}_{\varrho_{\gamma}}^\prime)\sqrt{\varsigma}),
\end{equation}

\noindent
where $\mathrm{L}_{\varsigma}^{-1} \chi= {\varsigma}^{-1}\chi$ is the operator of left multiplication by ${\varsigma}^{-1}$ isomorphic to the operator ${\varsigma}^{-1}\otimes I_{\bar{A}}$, and $\mathrm{L}_{\varrho}^\prime \chi= \chi \varrho$ is the right multiplication by $\varrho$ isomorphic to the operator $\mathcal{I}_{A} \otimes \tilde{\varrho}$, $g(r)$ is any strictly positive function $g(r)$ at $r\geq 1$ with $g(1)=0$ and operator-convex, $\lambda$ is a faithful, semi-finite trace.

Note that this quantity is introduced in [HB 2006] in order to characterize the "distance" as an information divergence between states.

Quantum relative entropy of Hammersley-Belavkin type (known as $\gamma$ type elsewhere) includes the other two relative quantum information as special cases.

\textbf{Corollary} If $\gamma = I$, it gives quantum relative entropy of Araki-Umegaki type (known as $a$ type elsewhere) to compound state $\varpi$ on the algebra $\mathcal{A} \otimes\mathcal{B}$, (or information divergence of the state $\varpi$ with respect to a reference state $\varphi$) defined by the density operator $\omega, \phi$ of these states $\varpi$ and $\varphi$ as
\begin{equation}
\mathcal{R}^{(a)}(\varpi; \varphi)=\mathrm{Tr}[\omega (\ln \omega-\ln \phi)].
\end{equation}

This quantity is used in most definitions of quantum relative entropy. However unlike the classical case, this is not only possible choice for informational divergence of the states $\varpi$ and $\varphi$, and it does not relate explicitly the informational divergence to the Radon-Nikodym (RN) density $\omega_{\varphi}=\phi^{-1/2}\omega \phi^{-1/2}$ of the state $\varpi$ with respect to $\varphi$ as in the classical case.

\textbf{Corollary} If $\gamma =\sigma$, it gives quantum relative entropy of Belavkin-Staszewski type (known as $b$ type elsewhere) introduced in [HB 2006] as
\begin{equation}
\mathcal{R}^{(b)}(\varpi; \varphi)=\mathrm{Tr}[\omega \ln (\phi ^{-1}\omega)],
\end{equation}

\noindent
where $\omega \ln (\phi ^{-1}\omega)=\ln (\omega\phi ^{-1})\omega $ is understood as the Hermitian operator
\begin{equation}
\omega^{1/2} \ln (\omega^{1/2}\phi ^{-1}\omega^{1/2})\omega^{1/2}.
\end{equation}

This relative entropy can be explicitly written in terms of the RN density $\omega_{\varphi}$ as $\mathcal{R}^{(b)}(\varpi; \varphi)=\varphi(r(\omega_{\varphi}))$, where $r(\omega_{\varphi})=\omega_{\varphi}\ln \omega_{\varphi}$.

\textbf{Remark} In finite dimensions and faithful states, the Belavkin-Staszewski information divergence based on quantum relative entropy of Belavkin-Staszewski type gives better distinction of $\omega$ and $\phi$ [OP 1993] in the sense that it is greater than quantum relative entropy of Araki-Umegaki type, and that it satisfies the following important monotonicity property.

\textbf{Remark} Quantum relative entropy of Hammersley-Belavkin type (known as $\gamma$ type elsewhere) is more generous quantum relative entropy, and includes Araki-Umegaki type (known as $a$ type elsewhere) and Belavkin-Staszewski  type (known as $b$ type elsewhere), but can not exhaust all possibilities expect for commutative algebra, for example, the trace distance $\mathcal{R}_{tr}(\varrho;\varsigma)=\lambda (|\varrho-\varsigma|)$ and the fidelity distance $\mathcal{R}_{fid}(\varrho;\varsigma)=1-\lambda (|\sqrt{\varrho}\sqrt{\varsigma}|)$ are not of Hammersley-Belavkin type (known as $\gamma$ type elsewhere)[HB 2006]. Undoubtedly, it is a challenge to obtain a general information-theoretic definition for quantum relative entropy which reduces to a general mathematical formula for quantum relative entropy.

For quantum relative entropy of the three types, it is easy to obtain the well-known monotonicity inequality.

\textbf{Theorem} Given a normal completely positive unital map $K:\mathcal{M}\rightarrow \mathcal{M}^0$, if $\varpi=\varpi_0K, \varphi=\varphi_0K$, then for both relative entropies,
\begin{equation}
\mathcal{R}(\varpi; \varphi)\leq \mathcal{R}(\varpi_0; \varphi_0).
\end{equation}

\textbf{Remark} In fact, this monotonicity property is proved in [Lindblad 1973, Holevo 1998a] for Araki-Umegaki type (known as $a$ type elsewhere). And Belavkin-Staszewski type (known as $b$ type elsewhere) and Hammersley-Belavkin type (known as $\gamma$ type elsewhere) satisfies a system of properties [HB 2006] for quantum relative entropy including this monotonicity inequality. Therefore this monotonicity property holds for the three quantum relative entropies.

\subsection{Quantum Estimate}
In a quantum estimate, we are given quantum observations distributed according to an unknown distribution $P_{\theta} \in X$ associated with award $Q$ (where $X$ is a predual space of Hermitian $W^*$-continuous functionals on a $W^*$-algebra $X^*=\mathcal{A}$), which Nature chooses randomly from the set $\{P_{\theta}: \theta \in \Theta \}$ according to a known prior distribution $\mu$ on $\Theta$, and we produce an estimate $M$ for the unknown distribution $P_{\theta}$. In the end, we will suffer a relative entropy cost $\mathcal{R}(P;M)$, measuring the quality of this estimate, thus the whole utility is taken as $P \cdot Q -\mathcal{R}(P; M)$.

There are several mathematical quantum relative entropies, mainly Araki-Umegaki type, Belavkin-Staszewski type, Hammersley-Belavkin type, and similarly three different mathematical forms of game theoretic utilities. Below we concentrate on quantum relative entropy of Araki-Umegaki type, since other two types of quantum relative entropy are just mathematical complex in Radon-Nikodym (RN) derivatives but with the same intrinsic method.

Applying the convex conjugate $\mathcal{R}_{M,\mu}^*(Q)$ of $\mathcal{R}(P;M)$ with respect to $P$, that is,
\begin{equation}
\mathcal{R}_{M,\mu}^*(Q)= \max_{P \geq 0, P \cdot I=\mu} [P \cdot Q- \mathcal{R}(P;M)],
\end{equation}

\noindent
the estimate problem is the following optimization problem.
\begin{equation}
\min_{M \geq 0, M \cdot I=1} \max_{P \geq 0, P \cdot I=\mu} [P \cdot Q- \mathcal{R}(P;M)]= \min_{M \geq 0, M \cdot I=1} \mathcal{R}_{M,\mu}^*(Q),
\end{equation}

\noindent
where the cost function $\mathcal{R}(P;M)$ here is taken as quantum relative entropy of Araki-Umegaki type.

\textbf{Remark} If taking function $\mathcal{R}_1(P)$ as follows.
$$
\mathcal{R}_1(P) = \left\{
\begin{array}{ll}
\mathcal{R}(P;M) \ \ \ \ P \cdot I=\mu; \\
\infty \ \ \ \ P \cdot I \neq 1. \\
\end{array}
\right.
$$

\noindent
Then we can rewrite $\mathcal{R}_{M,\mu}^*(Q)$ as follows.
\begin{equation}
\max_{P \geq 0} [P \cdot Q- \mathcal{R}_1(P)]=\mathcal{R}_{M,\mu}^*(Q),
\end{equation}

\noindent
where $\mathcal{R}_1(P)=\mathcal{R}(P;M)+ O_{A_1}(P)$, and the hyperplane $A_1=\{ P: P \cdot I=\mu\}$.

Applying the Lagrange Theorem, we will find.
\begin{equation}
\mathcal{R}_{M,\mu}^*(Q)=\max_{P \geq 0, P \cdot I=\mu} [P \cdot Q- \mathcal{R}_1(P)]=\min_{\lambda} \max_{P \geq 0} [P \cdot Q- \mathcal{R}_1(P)+\lambda (P \cdot I - \mu)].
\end{equation}

Since the function $P \cdot Q- \mathcal{R}_1(P)+\lambda (P \cdot I - \mu)$ is linear with respect to $\lambda$ and convex with respect to $P$, therefore the $\min$ and $\max$ can be exchanged, that is,
\begin{equation}
\min_{\lambda} \max_{P \geq 0} [P \cdot Q- \mathcal{R}_1(P)+\lambda (P \cdot I - \mu)]=\max_{P \geq 0} \min_{\lambda} [P \cdot Q- \mathcal{R}_1(P)+\lambda (P \cdot I - \mu)],
\end{equation}

\noindent
we can rewrite $\mathcal{R}_{M,\mu}^*(Q)$ as follows.
\begin{equation}
\mathcal{R}_{M,\mu}^*(Q)=\max_{P \geq 0, P \cdot I=\mu} [P \cdot Q- \mathcal{R}(P;M)]= \max_{P \geq 0} \min_{\lambda} [P \cdot Q- \mathcal{R}_1(P)+\lambda (P \cdot I - \mu)].
\end{equation}

\textbf{Remark} Considering this optimization problem at first with respect to $\lambda$, then the following formula is reached.
\begin{equation}
\mathcal{R}_{M,\mu}^*(Q)= \max_{P \geq 0}  [P \cdot Q- \mathcal{R}_1(P)+\min_{\lambda} \lambda (P \cdot I - \mu)]=\max_{P \geq 0} [P \cdot Q- \mathcal{R}(P;M)+O_{A_1}(P)].
\end{equation}

Applying the variational method to the function $U=P \cdot Q- \mathcal{R}_1(P)+\lambda (P \cdot I - \mu)$, we obtain the following results.
\begin{equation}
Q+ \ln M- \ln P+ (\lambda - 1)I=0,
\end{equation}
\begin{equation}
\mathrm{Tr} P= \mu.
\end{equation}

Simple calculations give us results as follows.
\begin{equation}
P^*=M \cdot \exp [Q +(\lambda -1)I],
\end{equation}
\begin{equation}
\mathcal{R}_{M,\mu}^*(Q)=P^* \cdot Q -\mathcal{R}_{M,\mu}(P^*)= - \mathrm{Tr} [M \exp (Q +\lambda -1) (\lambda -1)].
\end{equation}

Considering further about two conditions $P \cdot I = \mu$ and $M \cdot I = 1$, we will find following three results.
\begin{equation}
\exp (\lambda -1) = \frac{\mu}{\mathrm{Tr} [M \exp Q]},
\end{equation}
\begin{equation}
P^*=\mu \frac{M \cdot \exp Q}{ \mathrm{Tr} [M \exp Q]},
\end{equation}
\begin{equation}
\mathcal{R}_{M,\mu}^*(Q)=P^* \cdot Q -\mathcal{R}_{M,\mu}(P^*)= \mu \ln [\frac{1}{\mu}\mathrm{Tr} (M \exp Q)].
\end{equation}

\textbf{Remark} It is easy to see that $\mathcal{R}_{M,\mu}(P^*)=\mathcal{R}_{M,\mu}^*(Q)-P^* \cdot Q$ is the quantum relative entropy under the maximal utility over $P \geq 0$ and $P \cdot I=\mu$, or quantum relative capacity under given utility. It is one interesting problem to explain $\mathcal{R}_{M,\mu}^*(Q)= \mu \ln [\frac{1}{\mu} \mathrm{Tr} (M \exp Q)]$ completely in quantum information theory.

Easily we can find the following properties of $\mathcal{R}_{M,\mu}^*(Q)$.

\textbf{Lemma} The function $\mathcal{R}_{M,\mu}^*(Q)= \mu \ln [\frac{1}{\mu} \mathrm{Tr} (M \exp Q)]$ is monotonous with respect to $M$.

\textbf{Lemma} The function $\mathcal{R}_{M,\mu}^*(Q)= \mu \ln [\frac{1}{\mu}\mathrm{Tr} (M \exp Q)]$ is concave with respect to $M$.

Since the set $\{ M: M \geq 0, M \cdot I=1 \}$ is convex, there always exists the minimum of function $\mathcal{R}_{M,\mu}(P^*)$ with respect to $M$, and upon the above lemmas, we reach the following main result.

\textbf{Theorem} The minimax value, associated with the above problem, defined by
\begin{equation}
\overline{\mathrm{V}}= \inf_{M \geq 0, M \cdot I=1} \sup_{P \geq 0, P \cdot I=\mu} [P \cdot Q- \mathcal{R}(P;M)],
\end{equation}

\noindent
makes sense, and so does its minimax strategy.

\textbf{Remark} We can similarly define the maxmin value, associated with the above problem, by
\begin{equation}
\underline{\mathrm{V}}= \sup_{P \geq 0, P \cdot I=\mu} \inf_{M \geq 0, M \cdot I=1} [P \cdot Q- \mathcal{R}(P;M)],
\end{equation}

\noindent
but it is a problem if this maxmin value and its maxmin strategy exist and if this maxmin value is equal to the minimax value.

\subsection{Convex Conjugate View}
In the above analysis, we applied convex conjugate $\mathcal{R}_M^*(Q)$ of quantum relative entropy $\mathcal{R}_M(P)$, that is,
\begin{equation}
\mathcal{R}_{M,\mu}^*(Q)=\max_{P \geq 0, P \cdot I=\mu} \{P \cdot Q- \mathcal{R}_M(P)\},
\end{equation}

\noindent
and obtained the following formula
\begin{equation}
\mathcal{R}_{M,\mu}^*(Q)= \mu \ln [\frac{1}{\mu} \mathrm{Tr} (M \exp Q)],
\end{equation}

\noindent
but starting from the result $\mathcal{R}_{M,\mu}^*(Q)= \mu \ln [\frac{1}{\mu} \mathrm{Tr} (M \exp Q)]$, and applies the biconjugate property of convex conjugate, we can define the quantum relative entropy $\mathcal{R}_\mu(\rho; M)$ as follows.

\textbf{Definition} Quantum relative entropy $\mathcal{R}_\mu(\rho; M)$ of $\rho$ relative to $M$ is defined as
\begin{equation}
\mathcal{R}_\mu(\rho; M) \equiv \mathcal{R}_{M,\mu}(\rho)=\max_Q \{<\rho, Q> - \mathcal{R}_{M,\mu}^*(Q)\},
\end{equation}

\noindent
where $\mathcal{R}_{M,\mu}^*(Q)= \mu \ln [\frac{1}{\mu} \mathrm{Tr} (M \exp Q)]$.

In particular, for the case $\mu=1$,
\begin{equation}
\mathcal{R}_1(\rho; M) \equiv \mathcal{R}_{M,1}(\rho)=\max_Q \{<\rho, Q> - \mathcal{R}_{M}^*(Q)\},
\end{equation}

\noindent
where $\mathcal{R}_{M}^*(Q)= \ln [\mathrm{Tr} (M \exp Q)]$.

Obviously, the general mathematical form of this quantum relative entropy may not include the mathematical form of the Araki-Umegaki type, Belavkin-Staszewski type, or Hammersley-Belavkin type.

Here we obtain one property for this quantum relative entropy.

\textbf{Theorem}
\begin{equation}
\forall \mu > 0, \mathcal{R}_\mu(\rho; M) = \mu \mathcal{R}_1(\frac{1}{\mu}\rho; \frac{1}{\mu} M).
\end{equation}

Proof: According to the definition of quantum relative entropy,
\begin{equation}
\mathcal{R}_\mu(\rho; M) \equiv \mathcal{R}_{M,\mu}(\rho)
\end{equation}
\begin{equation}
=\max_Q \{<\rho, Q> - \mu \ln [\frac{1}{\mu}\mathrm{Tr} M \exp Q)]\}
\end{equation}
\begin{equation}
=\mu \max_Q \{<\frac{1}{\mu}\rho, Q> - \ln [\mathrm{Tr}(\frac{1}{\mu} M \exp Q)]\}
\end{equation}
\begin{equation}
=\mu \mathcal{R}_1(\frac{1}{\mu}\rho; \frac{1}{\mu} M).
\end{equation}

\subsection{Monotonicity of Quantum Relative Entropy}

To obtain its monotonicity of $\mathcal{R}_1(\rho; M)$ (or $\mathcal{R}(\rho; M)$ without notation confusion), we need several lemmas in our notation system.

Let $\rho, \zeta$ be positive trace class operators in a separable Hilbert space $\mathcal{H}$, $\Gamma$ a trace-preserving from $\mathcal{B}(\mathcal{H})$ to a von Neumann subalgebra $\mathcal{A}$.

\textbf{Lemma 1}([Kovacs 1966]) Let $\mathcal{A}$ be a finite von Neumann algebra. For each $X \in \mathcal{A}$ and each inner $^*-$automorphism $\alpha$ of $\mathcal{A}$, there exists $\overline{X} \in \mathcal{A}$ such that $\frac{1}{N}\sum_{n=0}^{N-1} \alpha^n (X) \longrightarrow \overline{X}$ as $N \longrightarrow \infty$ in the strong operator topology.

\textbf{Lemma 2} ([schwartz 1967]) Let $\mathcal{A}^\prime$ denote the group of all unitary transformations in $\mathcal{A}$,  $E(\mathcal{A}^\prime)$ the set of all non-negative real valued functions on $\mathcal{A}^\prime$ which vanish expect at a finite number of points of $\mathcal{A}^\prime$ and which satisfy
\begin{equation}
\sum_{U\in \mathcal{A}^\prime} f(U)=I.
\end{equation}

\noindent
Take $f(X)=\sum_{U\in \mathcal{A}^\prime} f(U)UXU^*$ for each bounded operator $X$, then for each bounded operator $X$, there exists a sequence $f_n \in E(\mathcal{A}^\prime)$ such that $f_n(X)$ converges weakly to an element of $\mathcal{A}^\prime$.

\textbf{Lemma 3} Let $X \in T(\mathcal{H})$ be a trace class operators in a separable Hilbert space $\mathcal{H}$, and $K(X)$ the weakly closed convex hull of the set
\begin{equation}
\{UXU^{-1}, \mathrm{unitary} U \in \mathcal{A}^\prime\}.
\end{equation}

\noindent
Then
\begin{equation}
K(X) \cap \mathcal{A} = \{\Gamma (X)\},
\end{equation}

\noindent
$\{\Gamma (Y)\}=\{\Gamma (X)\}$ for all $Y \in K(X)$.

Moreover, let $E(\mathcal{A}^\prime)$ be the set of nonnegative real functions on the set $U(\mathcal{A}^\prime)$ of unitary operators in $\mathcal{A}^\prime$ which are nonzero only on a finite number of points and which satisfy $\sum f(U)=I$. Take $f(X)=\sum f(U)UXU^{-1}$, then there is a sequence $\{f_n\} \subset E(\mathcal{A}^\prime)$ such that $f_n(X)\longrightarrow \Gamma(X)$.

Proof: $\Gamma$ is normal, since the trace is normal, hence ultra-weakly continuous.

If $X \in T(\mathcal{H})$ be a trace class operators in a separable Hilbert space $\mathcal{H}$, $\Gamma(X)$ is the unique element of $\mathcal{A}$ such that
\begin{equation}
\mathrm{Tr} (\Gamma(X) Y)= \mathrm{Tr}(XY),
\end{equation}

\noindent
for all $Y \in \mathcal{A}$, which implies that $\Gamma(UXU^{-1})=\Gamma(X)$ for all unitary $U \in \mathcal{A}^\prime$, hence $\Gamma$ is ultra-weakly continuous $\Gamma (X)= \Gamma (Y)$ for all $Y \in K(X)$.

The first statement results from Lemma 1; the last statement follows from Lemma 2.

\textbf{Lemma 4} $\mathcal{R}(\rho; \zeta)$ is jointly convex in $\rho$ and $\zeta$: If $\lambda_i > 0, \sum \lambda_i=1$, then
\begin{equation}
\mathcal{R}(\sum \lambda_i \rho_i; \sum \lambda_i\zeta_i) \leq \sum \lambda_i \mathcal{R}(\rho_i; \zeta_i).
\end{equation}

Proof: Since \begin{equation}
\mathcal{R}_\mu(\rho; \zeta) \equiv \mathcal{R}_{\zeta,\mu}(\rho)
\end{equation}
\begin{equation}
=\max_Q \{<\rho, Q> - \mu \ln [\frac{1}{\mu}\mathrm{Tr} (\zeta \exp Q)]\},
\end{equation}
\begin{equation}
\mathcal{R}(\sum \lambda_i \rho_i; \sum \lambda_i\zeta_i) = \max_Q \{<\sum \lambda_i \rho_i, Q> - \ln [\mathrm{Tr} (\sum \lambda_i\zeta_i \exp Q)]\}
\end{equation}
\begin{equation}
\leq \max_Q \sum \lambda_i \{<\rho_i, Q> - \ln [\mathrm{Tr} (\zeta_i \exp Q)]\}
\end{equation}

\noindent
by additivity of inner product, $<\sum \lambda_i \rho_i, Q>=\sum \lambda_i< \rho_i, Q>$, and convexity of logarithm function,
\begin{equation}
= \sum \lambda_i \max_Q  \{<\rho_i, Q> - \ln [\mathrm{Tr}(\zeta_i \exp Q)]\}
\end{equation}

\noindent
by the positivity of $\lambda_i$ for all $i$,
\begin{equation}
= \sum \lambda_i \mathcal{R}(\rho_i; \zeta_i)
\end{equation}
\noindent
by the definition of $\mathcal{R}(\rho_i; \zeta_i)$.

\textbf{Lemma 5} Let $P$ be a projection in $\mathcal{H}$, and take $\rho_P \equiv P \rho P$, etc., then
\begin{equation}
\mathcal{R}(\rho_P; \zeta_P) + \mathcal{R}(\rho_{I-P}; \zeta_{I-P}) \leq \mathcal{R}(\rho; \zeta).
\end{equation}

Proof: Note that $U = 2P-I$ is unitary and that
\begin{equation}
\rho^\prime \equiv \rho_P+ \rho_{I-P}=\frac{1}{2}(\rho+ U^+ \rho U).
\end{equation}

Following Lemma 4, we have
\begin{equation}
\mathcal{R}(\rho^\prime; \zeta^\prime) \leq \frac{1}{2}\mathcal{R}(\rho; \zeta) + \frac{1}{2}\mathcal{R}(U^+ \rho U; U^+ \zeta U ) = \mathcal{R}(\rho; \zeta).
\end{equation}

Note that
\begin{equation}
\mathcal{R}(\rho^\prime; \zeta^\prime) \geq \mathcal{R}(\rho_P; \zeta_P) + \mathcal{R}(\rho_{I-P}; \zeta_{I-P}),
\end{equation}

\noindent
we reach the result
\begin{equation}
\mathcal{R}(\rho_P; \zeta_P) + \mathcal{R}(\rho_{I-P}; \zeta_{I-P}) \leq \mathcal{R}(\rho; \zeta).
\end{equation}

\textbf{Lemma 6} Let ${P_n}$ be a sequence of projections such that $P_m \leq P_n$ for $m \leq n$, $\dim P_n$ is finite for all $n$, and $P_n\longrightarrow I$ strongly when $n \longrightarrow \infty$. Take $\rho_n = P_n \rho P_n$, then the sequences $\{\mathcal{R}(\rho_n; \zeta_n)\}$ are monotonously increasing and
\begin{equation}
\mathcal{R}(\rho_n; \zeta_n)\longrightarrow \mathcal{R}(\rho; \zeta).
\end{equation}

Proof: The monotonicity of $\{\mathcal{R}(\rho_n; \zeta_n)\}$ follows from Lemma 5.

Recalling the facts,
\begin{equation}
\mathrm{Tr}(P_n \rho^2) \longrightarrow \mathrm{Tr}\rho^2,
\end{equation}
\noindent
and
\begin{equation}
0 \leq \mathrm{Tr}[P_n (\rho_n^2 -\rho^2)]=\mathrm{Tr}[P_n \rho(I -P_n) \rho] \leq \mathrm{Tr}[\rho^2 (I -P_n)] \longrightarrow 0,
\end{equation}

\noindent
we have the result
\begin{equation}
\mathrm{Tr}(\rho-\rho_n)^2=\mathrm{Tr}(\rho^2-\rho_n^2)= \mathrm{Tr}[\rho^2 (I -P_n)] +\mathrm{Tr}[P_n (\rho_n^2 -\rho^2)] \longrightarrow 0,
\end{equation}

\noindent
but $\| \rho -\rho_n \|^2 \leq \mathrm{Tr}(\rho-\rho_n)^2$, consequently $\| \rho -\rho_n \|\longrightarrow 0$, that is, the convergence $\rho_n \longrightarrow \rho$ is uniform.

Note that $\mathcal{R}(\rho; \zeta)$ is lower semicontinuous under the convergence $(\rho_n; \zeta_n) \longrightarrow (\rho; \zeta)$,
\begin{equation}
\mathcal{R}(\rho; \zeta) \leq \lim \inf \mathcal{R}(\rho_n; \zeta_n),
\end{equation}

\noindent
but from Lemma 5, we know that $\mathcal{R}(\rho_n; \zeta_n) \leq \mathcal{R}(\rho; \zeta)$, hence we have
\begin{equation}
\lim \inf \mathcal{R}(\rho_n; \zeta_n) = \mathcal{R}(\rho; \zeta).
\end{equation}

\textbf{Lemma 7} Assume that $\{f_k\} \subset E(\mathcal{A}^\prime)$ satisfies
\begin{equation}
f_k (\rho) \longrightarrow \Gamma (\rho),
\end{equation}
\begin{equation}
f_k (\zeta) \longrightarrow \Gamma (\zeta)
\end{equation}

\noindent
weakly, then
\begin{equation}
\mathcal{R}(\Gamma (\rho); \Gamma (\zeta)) \leq \lim \inf \mathcal{R}(f_k (\rho); f_k (\zeta)) \leq \mathcal{R}(\rho; \zeta).
\end{equation}

Proof: Since $\Gamma$ is trace-preserving, use the spectral measure of $\Gamma (\rho)$ (where positive operator $\rho$ has the support projection $I$), there is a sequence of projections ${P_n}$ in $\mathcal{A}$ satisfying the conditions of Lemma 6.

From the definition of $\Gamma$, we have
\begin{equation}
\Gamma (P_n \rho P_n)= P_n \Gamma (\rho)P_n.
\end{equation}

Since $f_k$ is bulit up of elements of $\mathcal{A}^\prime$, we result that
\begin{equation}
f_k (P_n \rho P_n)= P_n \Gamma [f_k (\rho)]P_n.
\end{equation}

In the finite dimensional space, $\mathcal{H}_n = P_n \mathcal{H}$, $f_k (\rho_n) \longrightarrow \Gamma(\rho_n)$ is convergent uniformly, hence when $k \longrightarrow \infty$,
\begin{equation}
\mathcal{R}( f_k (\rho_n); f_k (\zeta_n)) \longrightarrow \mathcal{R}( \Gamma (\rho_n); \Gamma(\zeta_n)).
\end{equation}

According to Lemma 6, we obtain that
\begin{equation}
\mathcal{R}(\rho; \zeta)=\sup \mathcal{R}(\rho_n;\zeta_n),
\end{equation}

\noindent
hence $\mathcal{R}(\rho; \zeta)$ is lower semicontinuous, that is
\begin{equation}
\mathcal{R}(\Gamma(\rho); \Gamma(\zeta)) \leq \lim \inf \mathcal{R}(f_k(\rho); f_k(\zeta)).
\end{equation}

Following Lemma 4, we have
\begin{equation}
\mathcal{R}(f_k(\rho); f_k(\zeta)) \leq \sum f_k (U) \mathcal{R}(U \rho U^+; U \zeta U^+)
\end{equation}
\begin{equation}
=\mathcal{R}(\rho; \zeta)
\end{equation}

\noindent
by the unitary invariance of $\mathcal{R}$.

Hence,
\begin{equation}
\mathcal{R}(\Gamma(\rho);\Gamma(\zeta)) \leq \mathcal{R}(\rho; \zeta).
\end{equation}

\textbf{Remark} According to Lemma 7, it is hard to prove Monotonicity Theorem since we do not know if there is a sequence $\{f_k\} \subset E(\mathcal{A}^\prime)$ which implements $\Gamma$ on both $\rho$ and $\zeta$.

\textbf{Theorem} (Monotonicity) For trace-preserving expectations $\Gamma$ from $\mathcal{B}(\mathcal{H})$ to a von Neumann subalgebra $\mathcal{A}$, if $\rho$ and $\zeta$ are positive trace class operators in a separable Hilbert space $\mathcal{H}$, then
\begin{equation}
\mathcal{R}(\Gamma(\rho); \Gamma(\zeta)) \leq \mathcal{R}(\rho; \zeta).
\end{equation}

Proof: Choose a sequence of projections ${P_n} \in \mathcal{A}$, such that $P_m \leq P_n$ for $m \leq n$, $\dim P_n$ is finite for all $n$, and $P_n\longrightarrow I$ strongly when $n \longrightarrow \infty$, and $\{f_k\} \subset E(\mathcal{A}^\prime)$ such that $f_k (\rho) \longrightarrow \Gamma (\rho)$ weakly. Then
\begin{equation}
f_k (\rho_n) \longrightarrow \Gamma (\rho_n)
\end{equation}

\noindent
in norm.

For a given $k$, there exists $g_j \subset E(\mathcal{A}^\prime)$ such that $g_j [f_k (\zeta)] \longrightarrow \Gamma(\zeta)$ weakly when $j\longrightarrow \infty$, then
\begin{equation}
g_j [f_k (\zeta_n)] \longrightarrow \Gamma(\zeta_n)
\end{equation}

\noindent
in norm.

If $\|(f_k - \Gamma) \rho_n \| \leq \varepsilon (k)$, choose $g_{j,k}$ such that
\begin{equation}
\| (g_{j,k} f_k - \Gamma) (\zeta_n) \| \leq \varepsilon (k).
\end{equation}

Obviously
\begin{equation}
\| (g_{j,k} f_k  - \Gamma) (\rho_n) \|= \| g_{j,k} (f_k  - \Gamma) (\rho_n) \| \leq \|(f_k  -  \Gamma) (\rho_n) \|\leq \varepsilon (k).
\end{equation}

Then $h_k\equiv g_{j,k} f_k$ satisfies
\begin{equation}
h_k (\rho_n) \longrightarrow \Gamma (\rho_n),
\end{equation}
\begin{equation}
h_k (\zeta_n) \longrightarrow \Gamma (\zeta_n)
\end{equation}

\noindent
in norm.

According to the proof of Lemma 7, we have
\begin{equation}
\mathcal{R}(\Gamma(\rho_n); \Gamma(\zeta_n)) \leq \mathcal{R}(\rho_n; \zeta_n),
\end{equation}

\noindent
and further following Lemma 6,
\begin{equation}
\mathcal{R}(\Gamma(\rho); \Gamma(\zeta)) \leq \mathcal{R}(\rho; \zeta).
\end{equation}

\textbf{Corollary} Let ${P_k}$ be a set of mutually orthogonal projections in $\mathcal{H}$ satisfying $\sum P_k = I$ and the map $\Gamma: \rho \longrightarrow \sum P_k \rho P_k$ is trace-preserving describing the interaction of a finite quantum system with a classical apparatus measuring an observable with eigen-spaces $P_k$, then
\begin{equation}
\mathcal{R}(\Gamma(\rho); \Gamma(\zeta)) = \sum \mathcal{R}(P_k \rho P_k; P_k \zeta P_k) \leq \mathcal{R}(\rho; \zeta).
\end{equation}

\section{Quantum Mutual Entropy}
Quantum mutual entropy was discussed in [CA 1997] via entropy diagram, which seems not starting originally from the information-theoretic sense. In the view of quantum control, quantum mutual entropy is extensively researched, starting from [BS 2005] and more recently [Belavkin 2001a, 2001b, BO 2001, BO 2002, SW 2006]. Belavkin and Ohya [BO 2001, 2002] introduced quantum mutual information as the von Neumann negaentropy $\mathcal{R}(\varpi)=-\mathcal{S}(\varpi)$ of the entangled compound state related to negaentropy $\mathcal{R}(\varrho \otimes \varsigma)=-\mathcal{S}(\varrho \otimes \varsigma)$ of the product of marginal states, i.e. as the relative negaentropy $\mathcal{R}^{(a)}(\varpi; \varphi)=-\mathcal{S}^{(a)}(\varpi; \varphi)$, in the sense of Lindblad, Araki and Umegaki relative entropy [Lindblad 1973, Araki 1976, Umegaki 1962] with respect to $\varphi=\varrho \otimes \varsigma$.

Naturally this approach treats quantum mutual entropy via entanglement based on quantum relative entropy, therefore we can further define quantum mutual entropy via quantum entanglement.

\textbf{Definition} We define the quantum mutual information $\mathcal{I}_{\mathcal{A},\mathcal{B}}(\pi)=\mathcal{I}_{\mathcal{B},\mathcal{A}}(\pi^{\ast })$ of both types in a compound state $\omega $ achieved by a quantum entanglement $\pi :\mathcal{B}\rightarrow \mathcal{A}_{\ast }$, or by $\pi^{\ast}:\mathcal{A}\rightarrow \mathcal{B}_{\ast }$ with
\begin{equation}
\varrho (A)=\varpi (A\otimes I)=\mathrm{Tr}_{\mathcal{G}}[A\rho
],\varsigma
(B)=\varpi (I\otimes B)=Tr_{\mathcal{H}}[B\sigma]
\end{equation}

\noindent
as the quantum relative entropy of the state $\varpi$ on $\mathcal{M}=\mathcal{A}\otimes \mathcal{B}$ with the respect to the product state $\phi =\varrho \otimes \varsigma $:
\begin{equation}
\mathcal{I}_{\mathcal{A},\mathcal{B}}(\pi )=\mathcal{R}(\omega;\rho \otimes \sigma).
\end{equation}

\textbf{Theorem} Let $\lambda :\mathcal{B}\rightarrow \mathcal{A}_{\ast}^{0}$ be an entanglement of the state $\sigma (B)=\mathrm{Tr}[\lambda
(B)]$ to $(\mathcal{A}^{0},\rho ^{0})$ with $\mathcal{A}^{0}\subseteq \mathcal{L}(\mathcal{G}_{0})$, $\varrho ^{0}=\lambda (I)$ on $\mathcal{B}$, and
$\pi =\mathrm{K}_{\ast }\lambda $ be entanglement to the state $\rho =\rho^{0} \mathrm{K}$ on $\mathcal{A}\subseteq \mathcal{G}$ defined as the
composition of $\lambda $ with the predual operator $\mathrm{K}_{\ast}:\mathcal{A}_{\ast }^{0}\rightarrow \mathcal{A}_{\ast }$ normal completely positive unital map $\mathrm{K}:\mathcal{A}\rightarrow \mathcal{A}^{0}$, then the following monotonicity inequality holds
\begin{equation}
\mathcal{R}_{\mathcal{A},\mathcal{B}}(\pi )\leq \mathcal{R}_{\mathcal{A}^{0},\mathcal{B}}(\lambda ).
\end{equation}

Proof: This follows from the commutativity of the following diagrams:
\[
\begin{minipage}{2in}
\xymatrix{
\mathcal{A}_* &\mathcal{A}_*^0\ar[l]_{K_*}\\
&&\mathcal{B}\ar[ull]^\pi \ar[ul]_{\lambda}\\
}
Commutative diagram for entanglement $\pi$
\end{minipage}
\qquad
\begin{minipage}{2in}
\xymatrix{
\mathcal{A} \ar[r]^{K} \ar[drr]_{\pi_*}
&\mathcal{A}^0 \ar[dr]^{\lambda _*} \\
&&\mathcal{B}_*\\
}
Dual commutative diagram for entanglement $\pi_*$
\end{minipage}
\]

Applying the monotonicity property of our new quantum relative entropy on $\mathcal{M}=\mathcal{A}\otimes \mathcal{B}$ with respect to the predual map $\omega_{0}\mapsto (\mathrm{K}_{\ast }\otimes \mathrm{Id})(\omega _{0})$ corresponding to $\varpi_{0}\mapsto \varpi _{0}(\mathrm{K}\otimes\mathrm{Id
})$ as the ampliation $\mathrm{K}\otimes \mathrm{Id}$ of a normal completely positive unital map $\mathrm{K}:\mathcal{A}\rightarrow \mathcal{A}^{0}$.

\textbf{Definition} The maximal quantum mutual entropy $\mathcal{J}_{\tilde{\mathcal{B}},\mathcal{B}}(\pi_q)$ as the supremum
\begin{equation}
H_{\mathcal{B}}(\varsigma)=\sup_{\pi^*(I)=\sigma} \mathcal{I}_{\mathcal{B},\mathcal{A}}(\pi^*)=\mathcal{J}_{\mathcal{B},\tilde{\mathcal{B}}}(\pi_q^*)
\end{equation}

\noindent
over all entanglements $\pi^*$ of any $(\mathcal{A},\varrho)$ to $(\mathcal{B},\varsigma)$ is achieved on $\mathcal{A}^0=\tilde{\mathcal{B}}$, $\varrho^0=\tilde{\varsigma}$ by the standard quantum entanglement $\pi_q^*(A)=\sigma^{1/2}\tilde{A}\sigma^{1/2}$ for a fixed $\varsigma(B)=\mathrm{Tr}_{\mathcal{H}}[B\sigma]$ is named as entangled, or true quantum entropy of each type of the state $\varsigma$.

Similar definition for Araki-Umegaki type can be found in [Belavkin 2001a, 2001b; BO 2001, 2002], for Belavkin-Staszewski type in [BD 2007].

\textbf{Definition} We call the positive difference
\begin{equation}
H_{\mathcal{B}\mid \mathcal{A}}(\pi)=H_{\mathcal{B}}(\varsigma)- \mathcal{I}_{\mathcal{A},\mathcal{B}}(\pi)
\end{equation}

\noindent
entangled (or true quantum) conditional entropy respectively of each type on $\mathcal{B}$ with respect to $\mathcal{A}$.

Similar definition for Araki-Umegaki type can be found in [Belavkin 2001a, 2001b, 2002; BO 2001], for Belavkin-Staszewski type in [BD 2007].

\section{Quantum Communication Channel}

Entanglement-assisted quantum channel capacity, or entangled quantum channel capacity is extensively researched, for example, entangled quantum capacity [Belavkin 2001a, 2001b; BO 2001, 2002] via a common framework, entanglement-assisted quantum capacity [BSST 1999, 2001] via entangled quantum mutual entropy. We further discuss entangled quantum channel capacity via quantum mutual entropy upon our quantum relative entropy and its additivity property.

\subsection{Quantum Channel Capacity}

Let $\mathcal{B}\subseteq \mathcal{L}(\mathcal{H})$ be the $W^*$-algebra of operators in a (not necessarily finite dimensional unitary) Hilbert space $\mathcal{H}$. Generally we denote the set of states, i.e. positive unit trace operators in $\mathcal{B}(\mathcal{H})$ by $\mathcal{S}(\mathcal{H})$, the set of all $m$-dimensional projections by $\mathcal{P}_m(\mathcal{H})$ and the set of all projections by $\mathcal{P}(\mathcal{H})$.

\textbf{Definition} A quantum channel $\Gamma$ is a normal unital completely positive linear map (UCP) of $\mathcal{B}$ into the same or another algebra $\mathcal{B}^0 \subseteq \mathcal{B}(\mathcal{H}^0)$. These maps admit the Kraus decomposition, which is usually written in terms of the dual map $\Gamma^*:\mathcal{B}^0_*\rightarrow \mathcal{B}_*$ as $\Gamma^*(\sigma^0)=\sum_k A_k \sigma^0 A_k^*\equiv \Gamma_*(\sigma^0)$ ([Stinespring 1955], [Lindblad 1973], [Holevo 1998a]), $\Gamma(B)= \sum_k A^*_k B A_k$, for $A_k$ are operators $\mathcal{H}^0\rightarrow \mathcal{H}$ satisfying $\sum_k A^*_k A_k=I^0$.

For example, quantum noiseless channel in the case $\mathcal{B}=\mathcal{L}(\mathcal{H})$, $\mathcal{B}^0=\mathcal{L}(\mathcal{H}^0)$ is described by a single isometric operator $Y:\mathcal{H}^0 \rightarrow \mathcal{H}$ as $\Gamma(B)=Y^* BY$.  See for example [Holevo 1998a, Lindblad 1991] for the simple cases $\mathcal{B}=\mathcal{L}(\mathcal{H})$, $\dim (\mathcal{H})<\infty$.

A noisy quantum channel sends input pure states $\varsigma_0=\varrho_0$ on the algebra $\mathcal{B}^0=\mathcal{L}(\mathcal{H}^0)$ into mixed states described by the output densities $\sigma=\Gamma^*(\sigma^0)$ on $\mathcal{B}\subseteq \mathcal{L}(\mathcal{H})$ given by the predual $\Gamma_*=\Gamma^*\mid \mathcal{B}^0_*$ to the normal completely positive unital map $\Gamma:\mathcal{B}\rightarrow \mathcal{B}^0$ which can always be written as
\begin{equation}
\Gamma(B)=\mathrm{Tr}_{\mathcal{F}_+}[Y^\dag BY].
\end{equation}

\noindent
Here $Y$ is a linear operator from $\mathcal{H}^0\otimes \mathcal{F}_+$ to $\mathcal{H}$ with $\mathrm{Tr}_{\mathcal{F}_+}[Y^\dag Y]=I$, and ${\mathcal{F}_+}$ is a separable Hilbert space of quantum noise in the channel. Each input mixed state $\varsigma^0$ is transmitted into an output state $\varsigma=\varsigma^0 \Gamma$ given by the density operator
\begin{equation}
\Gamma^*(\varsigma^0)=Y(\sigma^0\otimes I_+)Y^\dag\in \mathcal{B}_*
\end{equation}

\noindent
for each density operator $\sigma^0\in \mathcal{B}_*^0$, the identity operator $I_+\in \mathcal{F}_+$.

We follow [Belavkin 2001a, 2001b; BO 2001, 2002] to denote $\mathcal{K}_q$ the set of all normal completely positive maps $\kappa :\mathcal{A} \rightarrow \mathcal{B}^0$ with any probe algebra $\mathcal{A}$, normalized as $\mathrm{Tr}\kappa (I)=1$, and $\mathcal{K}_q(\varsigma^0)$ be the subset of $\kappa\in \mathcal{K}_q$ with $\kappa (I)=\sigma^0$. We take the standard entanglement $\pi_q^0$ on $(\mathcal{B}^0,\varsigma^0)=(\mathcal{A}_0,\varrho^0)$, where $\varrho_0(A_0)=\mathrm{Tr}[A_0\rho_0]$ given by the density operator $\rho_0=\sigma^0$, and denote by $K$ a normal unital completely positive map $\mathcal{A}\rightarrow \mathcal{A}^0=\widetilde{\mathcal{A}}_0$ that decomposes $\kappa$ as $\kappa (A)=\rho_0^{1/2}\widetilde{K(A)}\rho_0^{1/2}$. It defines an input entanglement $\kappa^*=K_* \pi_q^0$ on the input of quantum channel as transpose-completely positive map on $\mathcal{A}_0=\mathcal{B}^0$ into $\mathcal{A}_*$ normalized to $\rho=K_*\rho^0$, $\rho^0=\widetilde{\rho}_0$.

The channel $\Gamma$ transmits this input entanglement as a true quantum encoding into the output entanglement $\pi=K_*\pi^0_q \Gamma\equiv K_*\lambda$ mapping $\mathcal{B}$ via the channel $\Gamma$ into $\mathcal{A}_*$ with $\pi(I)=\rho$. The mutual entangled information, transmitted via the channel for quantum encoding $\kappa$ is therefore $\mathcal{J}_{\mathcal{A},\mathcal{B}}(\kappa^*\Gamma)=\mathcal{J}_{\mathcal{A},\mathcal{B}}(K_*\pi^0_q \Gamma)=\mathcal{J}_{\mathcal{A},\mathcal{B}}(K_*\Gamma)$, where $\Gamma=\pi^0_q \Gamma$ is the standard input entanglement $\pi^0_q(B)=\sigma_0^{1/2}\tilde{B}\sigma_0^{1/2}$ with $\sigma_0=\widetilde{\sigma}^0$, transmitted via the channel $\Gamma$.

\textbf{Theorem} Given a quantum channel $\Gamma: \mathcal{B}\rightarrow \mathcal{B}^0$, and an input state $\varsigma^0$ on $\mathcal{B}^0$, the entangled input-output quantum information via a channel $\Gamma: \mathcal{B}\rightarrow \mathcal{B}^0$ achieves the maximal value
\begin{equation}
\mathcal{J}(\varsigma^0, \Gamma)=\sup_{\kappa\in \mathcal{K}_q(\varsigma^0)}\mathcal{J}(\kappa^*\Gamma)=\mathcal{I}_{\mathcal{A}^0, \mathcal{B}}(\Gamma),
\end{equation}

\noindent
where $\lambda=\pi_q^0 \Gamma$ is given by the corresponding extremal input entanglement $\pi_q^0$ mapping $\mathcal{B}^0=\tilde{\mathcal{A}^0}$ into $\mathcal{A}^0=\tilde{\mathcal{B}}^0$ with $\mathrm{Tr}[\pi_q(B)]=\varsigma^0(B)$ for all $B\in \mathcal{B}^0$.

Note that simliar theorem for Araki-Umegaki type can be found in [Belavkin 2001a, 2001b; BO 2001, 2002], for Belavkin-Staszewski type in [BD 2007].

The following definition depends on the commutativity of diagrams:
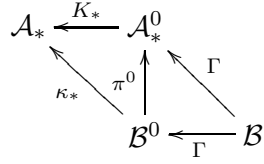
\[
\begin{minipage}{2in}
\xymatrix{
\mathcal{A}_*&\mathcal{A}_*^0\ar[l]_{K_*}\\
&\mathcal{B}^0\ar[ul]^{\kappa_*}\ar[u]^{\pi^0} &\mathcal{B}\ar[l]^{\Gamma}\ar[ul]_{\Gamma}\\
}
Commutative diagram for quantum channel $\Gamma$ \\ with standard entanglement $\pi^0=\pi^0_q$ for $\mathcal{A}=\widetilde{\mathcal{B}}^0$
\end{minipage}
\qquad
\begin{minipage}{2in}
\xymatrix{
\mathcal{A} \ar[r]^{K} \ar[dr]_{\kappa} &\mathcal{A}^0 \ar[d]^{\pi^0_*} \ar[dr]^{\lambda_*} \\
&\mathcal{B}_*^0 \ar[r]_{\Gamma_*} & \mathcal{B}_*\\
}
Dual commutative diagram for quantum channel $\Gamma$  \\with standard entanglement $\pi^0_*$ for $\mathcal{A}^0_*=\widetilde{\mathcal{B}}^0_*$
\end{minipage}
\]

\textbf{Definition} Given a quantum channel $\Gamma: \mathcal{B}\rightarrow \mathcal{B}^0$, and a input state $\varsigma^0$ on $\mathcal{B}^0$, we can define the input-output quantum entropy as the maximal quantum mutual information
\begin{equation}
\mathcal{J}(\varsigma^0, \Gamma)=\mathcal{I}_{\mathcal{B}^0, \mathcal{B}}(\pi_q^0 \Gamma)
\end{equation}

\noindent
for input standard entanglement of the state $\varsigma^0$ to the state $\varrho^0=\widetilde{\varsigma}^0$.

Note that similar definition for Araki-Umegaki type can be found in [Belavkin 2001a, 2001b; BO 2001, 2002], for Belavkin-Staszewski type in [BD 2007].

\subsection{Additivity of Quantum Channel Capacity}

Here and below for notational simplicity we implement the agreements $\mathcal{A}_0^i=\mathcal{B}_i^0$, $\varrho_0^i=\varsigma_i^0$, $\mathcal{A}_0^{\otimes}=\otimes_{i=1}^n \mathcal{B}_i^0$, $\varrho_0^{\otimes}=\otimes_{i=1}^n \varsigma_i^0$ such that $\varsigma_0^{\otimes}=\otimes_{i=1}^n \rho_i^0$ is transposed input state $\widetilde{\varrho}_0^{\otimes}=\otimes_{i=1}^n \widetilde{\varsigma}_i^0$ on $\mathcal{B}_0^{\otimes}=\otimes_{i=1}^n \mathcal{A}_i^0$ with $\widetilde{\mathcal{B}}_i^0=\mathcal{A}_i^0 \equiv \mathcal{B}_0^i=\widetilde{\mathcal{A}}_0^i$, $\widetilde{\varsigma}_i^0=\varrho_i^0 \equiv \varsigma_0^i=\widetilde{\varrho}_0^i$,.

Let $\Gamma_i$ be channels respectively from the algebra $\mathcal{B}_i$ on $\mathcal{H}_i$ to $\mathcal{B}_i^0$ on $\mathcal{H}_i^0$ for $i=1,2,...,n$, and let $\Gamma ^{\otimes}=\otimes_{i=1}^n \Gamma_i$ be their tensor product.

We have the additivity property of this entangled input-output quantum entropy upon the monotonicity property.

\textbf{Theorem} Let $\Gamma ^{\otimes}$ be product channel from the algebra $\mathcal{B} ^{\otimes}=\otimes_{i=1}^n \mathcal{B}_i$ to $\mathcal{A} _0^{\otimes}=\otimes_{i=1}^n \mathcal{A}^i_0$, and let $\varrho _0^{\otimes}=\otimes_{i=1}^n \varrho^i_0$ be the tensor product of input states $\varsigma^i_0$ on $\mathcal{B}_0^i$, then
\begin{equation}\label{eq154}
\mathcal{J}(\varrho_0 ^{\otimes}, \Gamma ^{\otimes})=\sum ^n_ {i=1} \mathcal{J}(\varrho_0^i, \Gamma_i).
\end{equation}

\textbf{Proof}: Take $\Gamma _{i*}:\mathcal{B}_{i*}^0 \rightarrow \mathcal{B}_{i*}$, and $\varrho _0^i \in \mathcal{B}_{i*}^0$, $\varsigma_i=\Gamma_{i*} (\varrho _0^i) \in \mathcal{B}_{i*}$, and $K^{(n)}_*:\mathcal{A}^{\otimes}_* \rightarrow \mathcal{A}^{(n)}_*$, where $\mathcal{A} ^{\otimes}_{0*}=\otimes_{i=1}^n \mathcal{B}_{i*}^0$, but $\mathcal{A}^{(n)}_*$ is predual to a general, not necessarily product algebra $\mathcal{A}^{(n)}\subseteq \mathcal{L}(\mathcal{G}^{(n)})$. For $\pi^{(n)}=K^{(n)}_*\pi_q^{0 \otimes}\Gamma^{\otimes}$, below we consider quantum mutual entropy $\mathcal{I}_{\mathcal{A}^{(n)},\mathcal{B}^{\otimes}}(\pi^{(n)})$ as quantum relative entropy
\begin{equation}
\mathcal{R}((K^{(n)}_* \otimes \Gamma ^{\otimes}_*)\widetilde{{\omega}}_0^{\otimes}: K^{(n)}_*(\sigma_0^{\otimes}){\otimes}\Gamma_*^{\otimes}({\varrho}_0^{\otimes})),
\end{equation}

\noindent
where $\widetilde{\omega}_0^{\otimes}=\otimes_{i=1}^n \widetilde{\omega}_0^i$ is the density operator of the standard compound state $\otimes_{i=1}^n \varpi_0^i$ with $\varpi_0^i(A_i\otimes B_i)=\omega_i^0(A_i\otimes B_i)=\mathrm{Tr}[B_i\sqrt{\rho_i^0} \widetilde{A}_i \sqrt{\rho_i^0}]$ for $A_i\in \widetilde{\mathcal{B}}_i^0, B_i \in \mathcal{B}_i^0$, corresponding to $\sigma^0_i=\rho^i_0$.

Applying monotonicity property of quantum relative entropy to the probe system $(\mathcal{G}^{(n)},\mathcal{A}^{(n)})$ for this given $\varrho _0^i$ and $\Gamma_i$, we obtain
\begin{equation}
\mathcal{R}((K^{(n)}_* \otimes \Gamma ^{\otimes}_*)\widetilde{\omega}_0^{\otimes};K^{(n)}_*(\sigma_0^{\otimes}){\otimes}\Gamma_*^{\otimes}({\rho}_0^{\otimes}))
\end{equation}
\begin{equation}
\leq \mathcal{R}((Id ^{\otimes} \otimes \Gamma^{\otimes})\widetilde{\omega}_0^{\otimes};Id ^{\otimes} (\sigma_0^{\otimes})\otimes \Gamma_*^{\otimes}({\rho}_0^{\otimes}))
\end{equation}
\begin{equation}
=\sum ^n_ {i=1} \mathcal{R}((Id \otimes \Gamma_{i*})(\widetilde{\omega} _0);Id (\sigma^i_0)\otimes \Gamma_{i*}(\rho^i_0)),
\end{equation}

\noindent
where $\sigma^i_0=\rho_i^0=\widetilde{\rho}^i_0$, $\rho^i_0=\sigma_i^0=\widetilde{\sigma}^i_0$.

The suprema over $K^{(n)}$ is achieved on $K^{(n)}=Id^{\otimes}$ identically mapping $\mathcal{A} ^{(n)}=\otimes_{i=1}^n \mathcal{A}_0^i$ to $\mathcal{B}_{0*} ^{\otimes}=\otimes_{i=1}^n \mathcal{B}_0^i$, where $\mathcal{B}_0^{i}=\widetilde{\mathcal{B}}_i^0$, coinciding with such $\mathcal{A} ^{(n)}$ due to $\mathcal{A}_0^{i}=\widetilde{\mathcal{B}}_i^0$.

Thus $\mathcal{J}(\varrho_0 ^{\otimes},\Gamma ^{\otimes})=\sum ^n_ {i=1} \mathcal{J}(\varrho_0^i,\Gamma_i)$.

\textbf{Definition} Given a normal unital completely positive map $\Gamma: \mathcal{B}\rightarrow \mathcal{A}$, the suprema
\begin{equation}
C_q(\Gamma)=\sup_{\kappa\in \mathcal{K}_q} \mathcal{I}_{\mathcal{A},\mathcal{B}}(\kappa^*\Gamma)=\sup_{\varsigma^0}\mathcal{J}(\varsigma^0,\Gamma)
\end{equation}

\noindent
is called the quantum channel capacity via entanglement, or q-capacity.

Let $\Gamma ^{\otimes}$ be product channel from the algebra $\mathcal{B} ^{\otimes}=\otimes_{i=1}^n \mathcal{B}_i$ to $\mathcal{A}_0 ^{\otimes}=\otimes_{i=1}^n \mathcal{B}_i^0$. The additivity problem for quantum channel capacity via entanglement is if it is true that
\begin{equation}
\mathcal{C}_q(\Gamma^{\otimes})=\sum ^n_ {i=1}\mathcal{C}_q(\Gamma_i).
\end{equation}

In the spirit of [Belavkin 2001a, 2001b; BO 2001, 2002; BD 2007], we prove this additivity property upon the monotonicity property of our quantum relative entropy.

\textbf{Theorem} Let $\Gamma ^{\otimes}$ be product channel from the algebra $\mathcal{B} ^{\otimes}=\otimes_{i=1}^n \mathcal{B}_i$ to $\mathcal{A}_0 ^{\otimes}=\otimes_{i=1}^n \mathcal{B}_i^0$, then
\begin{equation}
\mathcal{C}_q(\Gamma^{\otimes})=\sum ^n_ {i=1}\mathcal{C}_q(\Gamma_i).
\end{equation}

\textbf{Proof}: It simply follows from the additivity (\ref{eq154}). Indeed,
\begin{equation}
C_q(\Gamma^{\otimes})=\sup_{\kappa\in \mathcal{K}_q^{(n)}} \mathcal{I}_{\mathcal{A}^{(n)}, \mathcal{B}}(\kappa^*\Gamma^{\otimes})=\sup_{\varrho_0^{\otimes}}\mathcal{J}(\varrho_0^{\otimes},\Gamma^{\otimes})
=\sup_{\varrho_0^{\otimes}}\sum ^n_ {i=1}\mathcal{J}(\varrho_0^i,\Gamma_i)
\end{equation}

Therefore by further taking suprema over $\varrho_0^{\otimes}$ as over independently for each $i=1,2,...,n$, thus we have
\begin{equation}
\mathcal{C}_q(\Gamma^{\otimes})=\sum ^n_ {i=1}\sup_{\varrho_0^{\otimes}}\mathcal{J}(\varrho_0^i,\Gamma_i)=\sum ^n_ {i=1}\mathcal{C}_q(\Gamma_i),
\end{equation}

\noindent
which is the additivity property of entangled quantum channel capacity due to encodings via entanglement obviously.

\textbf{Remark} Note that there is no such additivity for Holevo channel capacity for a arbitrary channel $\Gamma: \mathcal{B}\rightarrow \mathcal{B}^0$. Indeed, this capacity is defined as the supremum
\begin{equation}
C_d(\Gamma)=\sup_{\kappa\in \mathcal{K}_d}\mathcal{I}_{\mathcal{A}, \mathcal{B}}(\kappa^*\Gamma)
\end{equation}

\noindent
over the smaller class $\mathcal{K}_d \subseteq \mathcal{K}_q$ of the diagonal (semiclassical) encodings $\kappa:\mathcal{A} \rightarrow \mathcal{B}_*^0$ corresponding to the Abelian algebra $\mathcal{A}$. This supremum cannot in general be achieved on the standard entanglement of $\mathcal{A}^0=\widetilde{\mathcal{B}}^0\equiv \mathcal{B}_0$ if $\mathcal{A}^0$ is non Abelian corresponding to the non Abelian input algebra $\mathcal{B}^0$. Therefore the supremum  $\mathcal{C}_d(\Gamma^{\otimes})\leq \sum ^n_ {i=1}\mathcal{C}_d(\Gamma_i)$ can be achieved not on a product Abelian algebra $\mathcal{A}^{(n)}$ as is was in the true quantum case where we could take $\mathcal{A}^{(n)}=\otimes^n_ {i=1}\mathcal{B}_0^i$ with non Abelian $\mathcal{B}_0^i=\widetilde{\mathcal{B}}_i^0$.

\section{Conclusion}
 This work so far is just a game theoretic-application to quantum information.

In the introduction to strategic game, we obtained a sufficient condition for minimax theorem, but it is still worth to explore the necessary condition of minimax theorem;

In the exploration of classical/quantum estimate, we found the existence of the minimax value of the game and its minimax strategy, but it is still interesting to discuss the existence of maximin value and its maximin strategy. One further problem is to give the general bounds on those values;

In the view of convex conjugate, we arrived at one approach to quantum relative entropy, quantum mutual entropy, and quantum channel capacity and discussed the monotonicity of quantum relative entropy and the additivity of quantum communication channel capacity, and it is still worth to explore other properties and their bounds.

\section{References}
[Borel 1953] E. Borel. The Theory of Play and Integral Equations with Skew Symmetric Kernels, Econometrica, 21, pp. 97-100, (1953).

[Kakutani 1941] S. Kakutani, A generalization of Brouwer's fixed point theorem, Duke Mathematical Journal 8, pp. 457-459, (1941).

[Nash 1950] J. F. Nash. Equilibrium Points in N-Person Games, Proceedings of the National Academy of Sciences of United States of America 36, pp. 48-49, (1950).

[Nash 1951] J. F. Nash. Non-Cooperative Games, Annals of Mathematics 54, pp. 286-295 (1951).

[NM 1944] J. von Neumann and O. Morgenstern. Theory of Games and Economic Behavior. Princeton University Press, (1944).

[Neumann 1928] J. von Neumann. On the Theory of Games of Strategy, Contributions to the Theory of Games, Volume 5, pp. 13-42 (1959)

[OR 1994] Martin J. Osborne, Ariel Rubinstein. A Course in Game Theory. MIT Press, (1994).

[Shannon 1948] C. E. Shannon. a mathematical theory of communication. The Bell System Technical Journal. Vol. 27. pp. 379-423, 623-656, Jul. Oct. (1948).

[Arnold 1989] Arnold, Vladimir Igorevich. Mathematical Methods of Classical Mechanics (second edition). Springer (1989).

[Rockafellar 1970] Rockafellar, Ralph Tyrell. Convex Analysis. Princeton University Press. (1970)

[Araki 1976] H. Araki, Relative Entropy of states of von Neumann Algebras, Publications RIMS, Kyoto University, 11,809 (1976)

[Belavkin 2001a] V. P. Belavkin, On Entangled Quantum Capacity. In: Quantum Communication, Computing, and Measurement 3. Kluwer/Plenum, pp. 325-333 (2001).

[Belavkin 2001b] V. P. Belavkin, On Entangled Information and Quantum Capacity, Open Sys. and Information Dyn, 8: pp. 1-18, (2001).

[BO 2001] V. P. Belavkin, M. Ohya, Quantum Entropy and Information in Discrete Entangled States, Infinite Dimensional Analysis, Quantum Probability and Related Topics 4  No. 2, pp. 137-160 (2001).

[BO 2002] V. P. Belavkin, M. Ohya, Entanglement, Quantum Entropy and Mutual Information, Proc. R. Soc. Lond. A 458  No. 2, pp. 209 - 231 (2002).

[BD 2007] V. P. Belavkin, X. Dai. Additivity of Entangled Channel Capacity given Quantum Input State. To appear in QPIC 2006, quant-ph/0702098

[BS 2005] V. P. Belavkin, R. L. Stratonovich, Optimization of Quantum Information Processing Maximizing Mutual Information, Radio Eng. Electron. Phys., 19 (9), pp. 1349, (1973).

[BSST 1999] C. H. Bennett, P. W. Shor, J. A. Smolin and A. V. Thapliyal, Entanglement assisted classical capacity of noisy quantum channels, Phys. Rev. Lett., vol. 83, pp. 3081-3084, (1999).

[BSST 2001] C. H. Bennett, P. W. Shor, J. A. Smolin and A. V. Thapliyal, Entanglement-Assisted Capacity of a Quantum Channel and the Reverse Shannon Theorem, quant-ph/0106052.

[CA 1997] N. Cerf and G. Adami, Von Neumann capacity of noisy quantum channels, Phys. Rev. A 56, pp. 3470-3483 (1997).

[HB 2006] S. J. Hammersley and V. P. Belavkin, Information Divergence for Quantum Channels, Infinite Dimensional Analysis.In:  Quantum Information and Computing. World Scientific, Quantum Probability and White Noise Analysis, VXIX pp. 149-166 (2006).

[Holevo 1998a] A. S. Holevo. Quantum coding theorems. Russian Math. Surveys 53:6 pp. 1295-1331, (1998)

[HOW 2005] Michal Horodecki, Jonathan Oppenheim, and Andreas Winter. Partial quantum information. Nature 436:pp. 673-676 (2005).

[Kovacs 1966] J. Kovacs and J. Szcs, Ergodic type theorems in von Neumann algebras, Acta Sci. Math. Szeged, 27, pp. 233-246 (1966).

[Lindblad 1973] G. Lindblad,  Entropy, Information and Quantum Measurements, Comm. in Math. Phys. 33, pp.305-322 (1973)

[Lindblad 1991] G. Lindblad, Quantum entropy and quantum measurements, in: Proc. Int. Conf. on Quantum Communication and Measurement, ed. by C. Benjaballah, O. Hirota, S. Reynaud, Lect. Notes Phys. 378, pp. 71-80, Springer-Verlag, Berlin (1991).

[Nakamura 1961] M. Nakamura and H. Umegaki, A note on entropy for operator algebras, Proc. Japan Acad. 37 , pp. 149-154 (1961).

[NC 2000] Nielsen, Michael A. and Isaac L. Chuang. Quantum Computation and Quantum Information. Cambridge University Press. 2000.

[OP 1993] M. Ohya and D. Petz, Quantum Entropy and its Use, Springer-Verlag, Berlin, (1993).

[schwartz 1967] Schwartz, Jacob T.: W*-algebras, New York: Gordon and Breach (1967)

[SW 2006] Benjamin Schumacher, and Michael D. Westmorel, Quantum mutual information and the one-time pad, quant-ph/0604207.

[Segal 1960] Segal, I.E.: A note on the concept of entropy, J. Math. Mech. 9, pp. 623-629 (1960)

[Stinespring 1955] W. F. Stinespring, Positive functions on C*-algebras, Proc. Amer. Math. Soc. 6, pp. 211-216 (1955).

[Umegaki 1962] H. Umegaki,Conditional expectation in an operator algebra. IV. Entropy and information,  Kodai Math. Sem. Rep. Volume 14, Number 2 , pp. 59-85(1962).

[Wehrl 1978] A. Wehrl, General properties of entropy, Rev. Mod. Phys., 50: pp. 221, (1978).

\end{document}